\documentclass[12pt]{article}
\usepackage{amsthm}
\usepackage[colorlinks=true,linkcolor=red,citecolor=blue]{hyperref}
\usepackage{rotating}
\usepackage{xcolor}
\usepackage{graphicx}
\usepackage{subfig}
\usepackage{amsmath}
\usepackage{amssymb}
\usepackage{enumitem}
\usepackage{natbib}
\usepackage{url} 
\usepackage{booktabs}
\usepackage{caption}
\usepackage{verbatim}
\usepackage{multirow}
\usepackage{comment}
\usepackage[linesnumbered,ruled]{algorithm2e}
\usepackage{bm}
\usepackage{mathtools}
\usepackage{setspace}
\allowdisplaybreaks

\newcommand{\blind}{1}
\begin{document}
	\def\spacingset#1{\renewcommand{\baselinestretch}%
{#1}\small\normalsize} \spacingset{1}
	
	\if1\blind
	{
		\title{\bf A Tree-based Semi-Varying Coefficient Model  for the COM-Poisson Distribution}
		\author{ Suneel Babu Chatla\thanks{The authors thank the associate editor and the three anonymous reviewers for their valuable suggestions which led to significant improvements in the  manuscript. This research is partially supported by Ministry of Science and Technology, Taiwan, grant 105-2410-H-007-034-MY3 (both authors) and grant 107-2811-M-007-1047 (first author). } \\
		{\small Department of Mathematical Sciences}\\
       {\small The University of Texas at El Paso}\\
     {\small El Paso, TX}\\
	Galit Shmueli \\
			{\small Institute of Service Science} \\
		{\small National Tsing Hua University} \\
        {\small Hsinchu, Taiwan} 
        }
        
		\maketitle
	} \fi
	
	\if0\blind
	{
		\bigskip
		\bigskip
		\bigskip
		\begin{center}
			{\LARGE\bf A Tree-based Semi-Varying Coefficient Model for the COM-Poisson Distribution}
		\end{center}
		\medskip
	} \fi
	

\begin{abstract}
  We propose a tree-based semi-varying coefficient model for the
  Conway-Maxwell-Poisson (CMP or COM-Poisson) distribution which is a two-parameter generalization of the Poisson distribution and is flexible enough to
  capture both under-dispersion and over-dispersion in count data. The
  advantage of  tree-based methods
  is their scalability to high-dimensional data.  
  We develop CMPMOB, an estimation procedure for 
  a semi-varying coefficient model, using model-based recursive partitioning (MOB). The proposed framework is broader than the existing MOB framework as it allows  node-invariant effects to be included in the model. To simplify the computational burden of the exhaustive search  employed in the original MOB algorithm,  a new split point estimation procedure is proposed by borrowing tools from change point estimation methodology. 
  The proposed 
  method uses only the estimated score functions without fitting models for each split point and, therefore, is computationally simpler. Since the tree-based methods only provide  a piece-wise constant approximation to the underlying smooth function, we propose the CMPBoost semi-varying coefficient model which uses the gradient boosting
   procedure 
   for estimation. 
   The usefulness of  the proposed methods are illustrated  using 
   simulation studies and 
   a real example from a bike sharing system in Washington, DC. 
\end{abstract}
\emph{Key words:} count data, gradient boosting, change point, model based recursive
partitioning, high-dimensional, 
bikesharing

\newpage
\spacingset{1.45} 
\section{Introduction}
	\label{sec:intro}

  Parametric models are often too restrictive for capturing complicated nonlinear
  relationships.  While one could use nonparametric models
  to provide more flexibility, 
  they often suffer  from the ``curse of dimensionality" \citep{park2015varying}. A  remedy 
  is to use structural nonparametric models. Two popular and useful examples of structural nonparametric models  are \emph{additive  models} proposed by \cite{breiman1985estimating} and \emph{varying coefficient models} proposed by \cite{hastie1993varying}. Since  additive models can be seen as a  special case of varying coefficient models, we focus on the latter.

  The  varying coefficient model is similar to an  ordinary regression model but the regression
  coefficients are allowed to vary based on the  value of other covariates \citep{hastie1993varying,park2015varying}, which are  often called \emph{moderator variables}. For a random sample of $n$, i.i.d. observations
  $\{y_i,\bm{x}_i^T,\bm{z}_i^T\}_{i=1}^n$ with
  $\bm{x}_i^T=(x_{i1},\ldots,x_{ip})$ and $\bm{z}_i^T=(z_{i1},\ldots,z_{iL})$, the varying coefficient model takes the
  following form:
  \begin{align}
    y_i &= \beta_0(\bm{z}_i)+x_{i1}\beta_1(\bm{z}_i)+\ldots + x_{ip}\beta_p(\bm{z}_i)+ \epsilon_i,
          \label{eqn:vcf-1}
    \end{align}
    where $\beta_k(\cdot)$, $0 \le k \le p$, are  $L-$dimensional functions
    and   the random errors ($\epsilon_i, i=1,\ldots,n$) are assumed to have mean zero and finite second moment. The special case of (\ref{eqn:vcf-1}) with $x_{ij}=1, j=1,\ldots,p$ is 
    the additive model. As long as the dimension of the moderator variables $\bm{z}_i$ $(L)$ is small, 
    Model (\ref{eqn:vcf-1}) can be estimated using  standard techniques such
    as smoothing splines \citep{,hastie1993varying}, penalized splines
    \citep{,woo07} and kernel smoothing \citep{,fan2008statistical}. However,
    most of these methods 
    break down when the dimension of $\bm{z}_i$ $(L)$ is large due to the curse of dimensionality. 
To overcome this problem, \citet{,wang2014boosted} proposed a novel 
    tree-based varying coefficient model estimation methodology 
    that is scalable to high-dimensional data. 

    In practice, not every 
    covariate in (\ref{eqn:vcf-1}) is assumed to 
    vary with $\bm{z}_i$. For example, interaction models \citep{,kaufman2018interaction} have main effects in addition to the interaction terms. Moreover, there are some covariates, such 
    as seasonal dummies in time series, that are better modeled as fixed effects rather than varying effects unless shown to actually vary. 
    Now, a question of interest is whether the coefficient $\beta_j(\bm{z}_i)$, $1 \le j\le p$, is really varying  \citep[e.g.][]{fan2000simultaneous}. This is equivalent to testing if the entire function is zero or constant, namely, $H_0:\beta_j(\bm{z}_i)=\beta_j$. Under $H_0$, Model (\ref{eqn:vcf-1}) is called a \emph{semi-varying coefficient model}  \citep[see e.g.][]{xia2004efficient,zhang2002local}. If $H_0$ is true, 
    estimating (\ref{eqn:vcf-1}) with its $p$ functions 
    yields estimates with inflated  
    variances. 
    
    In this study, we 
    develop a tree-based semi-varying coefficient model
    for  count data that is more flexible than (\ref{eqn:vcf-1}).  While the methodology proposed by \citet{,wang2014boosted} can
    be easily extended to  popular count  models such as Poisson
    regression or negative binomial regression, these models are not sufficiently flexible to     model  both over-dispersion and under-dispersion in 
    count data. For this reason, in this study, we consider the
    Conway-Maxwell-Poisson (CMP or COM-Poisson) distribution \citep{,shm05}
    which is a two-parameter generalization of the Poisson distribution that can capture both under-dispersion and over-dispersion. The
    estimation methodology for the CMP semi-varying coefficient model with the low-dimensional moderator variables (or $L$ is small) is developed by
    \citet{,chatla2017efficient}. In this study, we extend the estimation
    methodology to    high-dimensional moderator variables (or $L$ is large) using a tree-based estimation approach.  
  

The two most important choices in any tree building algorithm are the selection
of the splitting variable and the splitting value (split point). There exist different tree algorithms
 such as
GUIDE \citep{,loh2002regression}, CRUISE \citep{,kim2001classification} and
LOTUS \citep{,chan2004lotus} that use different methods to identify splitting variables and split points.
Although any tree building algorithm can be used, we
prefer the model based (MOB) recursive partitioning algorithm \citep{,zei08} as it uses
 coefficient-constancy tests to identify the splitting variables and is easy
to implement. However, the MOB algorithm  still uses an exhaustive search to estimate the optimal split point, which is computationally intensive. This is especially problematic for the CMP distribution, for which estimation 
is computationally heavier than 
for other 
generalized linear models such as Poisson and logistic regression. 

This work offers two contributions. The first contribution is that we 
develop an estimation methodology for 
semi-varying coefficient models (CMPMOB). The proposed procedure generalizes the MOB algorithm to allow global (node invariant) variables.
Additionally, to alleviate the computational burden of the existing exhaustive search method, we propose a new split point estimation method
that is less computationally intensive. 
The second contribution is a boosted estimation approach (CMPBoost), based on the gradient boosting algorithm, that provides smoother approximations of the underlying functions. 
CMPBoost is more flexible than the CMPMOB tree as it allows different sets of moderators for each parameter in the CMP distribution.

The remainder of the paper is organized as follows. Section \ref{sec:back} provides a brief introduction of  CMP regression and its current estimation methods. 
Section \ref{sec:mob-method} introduces the methodology and  algorithm for estimating the CMPMOB model. The performance of the proposed CMPMOB  is evaluated  using extensive simulations. 
In Section \ref{sec:gb}, we first provide a brief introduction of the existing
gradient boosting approach 
for the varying coefficient model and
then propose the CMPBoost estimation procedure. 
The usefulness of the proposed CMPBoost is evaluated using extensive simulations. 
Section
\ref{sec:mob-bikeshare} illustrates the pratical usefulness of the proposed CMPMOB and CMPBoost models 
by applying them to a real example from a bike
sharing application. 
We summarize our findings and conclusions 
in Section \ref{sec:mob-summary}.
 
 \section{Background}\label{sec:back}
 
 The CMP distribution is a member of  the two-parameter exponential family.  Suppose $y_i \sim CMP(\lambda_i,\nu_i)$, then the  probability mass function can be defined as
\begin{align*}
P(y_i = y)  = \frac{\lambda_i^y}{(y!)^{\nu_i} \zeta (\lambda_i, \nu_i)},  \quad \text{where} \quad \zeta(\lambda_i, \nu_i) = \sum_{s=0}^{\infty} \frac{\lambda_i^s}{(s!)^{\nu_i}}
\end{align*}
for  parameters $\lambda_i >0,\nu_i >0$ and $0 < \lambda_i <1 , \nu_i=0 $. Although $\lambda_i$ and $\nu_i$ are not the mean or the variance, it is useful to model both parameters due to the advantages of the
canonical link function \citep{,sel08}. Recently \citet{,huang2017mean} proposed an alternative parameterization (introducing a mean $\mu$ parameter) for the CMP distribution. \citet{,huang2017mean} models the mean $(\mu)$ parameter directly by additionally assuming a log link function. While this parameterization makes model interpretations easier by providing the model parameters in the scale of the mean, we did not consider it in this study because the existing estimation method for additive models (GAM) in \citet{,chatla2017efficient} was developed using the original  $(\lambda, \nu)$ parameterization. However, we note that the results remain similar whether one uses the original formulation $(\lambda, \nu)$ or the mean formulation $(\mu, \nu)$. The proposed methodologies in this study can be extended to the \citet{,huang2017mean} parameterization without much difficulty by modifying the estimation algorithms in \citet{,chatla2017efficient} appropriately. 

Assume that we have a random sample of $n$ observations $\{y_i,\boldsymbol{x}_i^T,\boldsymbol{w}_i^T \}_{i=1}^n $, where $\boldsymbol{x}_i^T=[1, x_{i1}, \cdots, x_{ip}]$ and $\boldsymbol{w}_i^T=[1, w_{i1}, \cdots, w_{iq}]$. 
We  denote the conditional mean and the conditional variance functions as $E[\cdot]$ and $V[\cdot]$ respectively. 
CMP regression is formulated as 
\begin{align}
\begin{split}
\ln(\lambda_i)= \eta_{1i} = \beta_0 + x_{i1}\beta_1+ \ldots + x_{ip}\beta_p, \\
\ln(\nu_i)= \eta_{2i} = \gamma_0 + w_{i1}\gamma_1+ \ldots + w_{iq}\gamma_q,
\label{eqn:cmp-reg-intro}
\end{split}
\end{align}
where $\bm{\beta}=(\beta_0, \beta_1, \ldots,\beta_p)^T \in \mathbb{R}^{p+1}$ and $\bm{\gamma}=(\gamma_0,\gamma_1,\ldots,\gamma_q)^T \in \mathbb{R}^{q+1}$.

The  log link is used for the  $\lambda_i$ model in (\ref{eqn:cmp-reg-intro}). As mentioned in \citet{sel08}, this choice of log link is useful for two reasons. First, it coincides with the link function in two well-known cases: in Poisson regression it reduces to $E[y_i] = \lambda_i$, while in logistic regression where $p_i =\frac{\lambda_i}{1+\lambda_i}$, it
reduces to $logit(p_i) = \ln \lambda_i$. The second advantage of using a log link function is that it leads to elegant estimation, inference, and diagnostics. At the same time,  a log link function is also considered for the $\nu_i$ model, although the canonical link is identity in order to restrict model predictions to the range $(0, \infty)$. This is important because, while $\gamma_i$ is unconstrained, $\nu_i$ $(i=1,\ldots,n)$ can only take positive values. 

In applications, it is common to treat $\nu_i$ as a nuisance parameter. For this reason, usually the data $\bm{w}_i^T$ contain only the intercept. Yet, since the $\nu_i$ parameter models the dispersion, it is always better to include covariates that can potentially control for it \citep{sel13}. In theory, one could use the same predictors for modeling both parameters. However, in practice, to avoid  collinearity issues, it is better to have at least one different covariate in either the $\ln(\lambda_i)$ or the $\ln(\nu_i)$ model.

 As mentioned in \citet{sel08}, the interpretation of the regression coefficients $\bm{\beta}$ and $\bm{\gamma}$ in (\ref{eqn:cmp-reg-intro}) is not straightforward. It is not possible to compare the conditional means directly as the relationship between the conditional mean  and the predictors is neither additive nor multiplicative. However, the result $E[y^{\nu_i}]=\lambda_i$ helps to provide a crude approximation for the relationship between regression coefficients and the conditional mean; one can simply divide the regression coefficients by $\nu_i$ to obtain the same scale as the conditional mean. We note that the parameterization in \citet{,huang2017mean} does not have this problem as the regression parameters are in the same scale of the conditional mean. However, the  model fit and statistical significance of regression coefficients remains the  same across both the parameterizations.

Using the model formulation in (\ref{eqn:cmp-reg-intro}), the log likelihood for 
observation $i$ can be written as 
\begin{align}
\ell_i(y_i, \boldsymbol{\beta}, \boldsymbol{\gamma} ) &=   y_i \boldsymbol{x}_i^T\boldsymbol{\beta}-\ln(y_i!) \exp\{\boldsymbol{w}_i^T\boldsymbol{\gamma}\}- \ln  \zeta_i (\exp\{\boldsymbol{x}_i^T\boldsymbol{\beta}\},\exp\{\boldsymbol{w}_i^T\boldsymbol{\gamma}\}) ,
\label{eqn:cmp-lk}
\end{align}
which yields the following score equations:
\begin{align}
\begin{split}
\frac{\partial \ell_i}{\partial \boldsymbol{\beta}^T} &= \boldsymbol{x}_i\left(y_i-\frac{\partial \ln \zeta_i} {\partial \ln \lambda_i}\right) =\boldsymbol{x}_i(y_i-E[y_i]),\\
\frac{\partial \ell_i}{\partial \boldsymbol{\gamma}^T} &= \boldsymbol{w}_i \bigg[\left(-\ln(y_i!)-\frac{\partial \ln \zeta_i} {\partial  \nu_i}\right)\nu_i \bigg]=\nu_i\boldsymbol{w}_i \bigg[\left(-\ln(y_i!)+E[\ln(y_i!)]\right) \bigg].
\end{split}
\label{eqn:score-intro}
\end{align}
The score equations in (\ref{eqn:score-intro}) are nonlinear as well as complicated due to involving 
derivatives of the infinite series $\zeta_i$ and cannot be solved directly. \citet{,chatla2017efficient} proposed a two-step iteratively reweighted least squares (IRLS) algorithm that uses the following normal equations to update both $\bm{\beta}$ and $\bm{\gamma}$ at each iteration to solve (\ref{eqn:score-intro}):
\begin{align}
\label{eq:wls1}
\sum_{i=1}^n \widehat{V}[y_i]\bm{x}_i\bm{x}_i^T\boldsymbol{\beta}^{(new)}&=\sum_{i=1}^n \widehat{V}[y_i] \widehat{t}_{i,1}\bm{x}_i, \\
\sum_{i=1}^n \widehat{V}[\ln y_i!] \widehat{\nu}_i^2\bm{w}_i\bm{w}_i^T\boldsymbol{\gamma}^{(new)}&=\sum_{i=1}^n \widehat{V}[\ln y_i!] \widehat{t}_{i,2} \widehat{\nu}_i\bm{w}_i,
\label{eq:wls2}
\end{align}
where $\widehat{V}[y_i]=\widehat{\text{Var}(y_i)}$, $\widehat{V}[\ln y_i!]=\widehat{\text{Var}(\ln y!)}$, $\widehat{t}_{i,1}=\bm{x}_i^T\widehat{\bm{\beta}}^{(old)}+ \widehat{V}^{-1}[y_i](y_i-\widehat{E[y_i]})$ and $\widehat{t}_{i,2}=\nu_i\bm{w}_i^T\widehat{\bm{\gamma}}^{(old)}+ \widehat{V}^{-1}[\ln y_i!](\widehat{E[\ln y_i!]}-\ln y_i!)$. 
The extensions to generalized additive models using penalized splines (using the Penalized IRLS algorithm) are also discussed in \citet{,chatla2017efficient}. We adopt these estimation methodologies in our proposed CMPMOB and CMPBoost approaches. 

\section{The CMPMOB Semi-Varying Coefficient Model} \label{sec:mob-method}

Consider a random sample of $n$ i.i.d. 
observations $\{y_i,\boldsymbol{x}_i^T,\boldsymbol{z}_i^T, \boldsymbol{w}_i^T \}_{i=1}^n $, where $\boldsymbol{x}_i^T=[x_{i1}, \cdots, x_{iP}]$, $\bm{z}_i^T=[z_{i1}, \cdots, z_{iL}]$ and $\bm{w}_i^T=[1, w_{i1}, \ldots, w_{iQ}]$. 
Using a regression formulation, the CMP semi-varying coefficient model (CMPMOB) takes the following form:
\begin{eqnarray}
\label{eqn:vc1}
\ln(\lambda_i)= \eta_{1i} = \bm{x}_{1i}^{T}\bm{\beta}(\bm{z}_{i})+ \bm{x}_{2i}^{T} \bm{\phi}_1, \\
\ln(\nu_i)= \eta_{2i} = \bm{w}_{1i}^{T}\bm{\gamma}(\bm{z}_{i})+ \bm{w}_{2i}^{T}\bm{\phi}_2,
\label{eqn:vc2}
\end{eqnarray} 
where $\bm{x}_{i}^{T}=(\bm{x}_{1i}^{T},\bm{x}_{2i}^{T})$,
$\bm{w}_{i}^{T}=(\bm{w}_{1i}^{T},\bm{w}_{2i}^{T})$, 
$\bm{\beta}(\cdot)$, $\bm{\gamma}(\cdot)$ are varying coefficients, and
$\bm{\phi}_1$, $\bm{\phi}_2$  are the node invariant (global) parameter vectors. 
We include node invariant coefficients in the model  because in practice the model should have some main effects in addition to the interaction effects modelled by the varying coefficients.  The node
invariant coefficients in both models can be parametric, 
nonparametric, or a combination. 
If we restrict the smooth functions to belong to a finite dimensional subspace,
then we can always represent the smooth functions as $\bm{x}_{2i}^T\bm{\phi}_1$
or $\bm{w}_{2i}^T\bm{\phi}_2$. While 
some overlap between $\bm{x}_i$, $\bm{w}_i$ and $\bm{z}_i$ is allowed, the degree of overlap is subjective and 
data-dependent. In practice, to avoid  multicollinearity issues and convergence challenges 
of the IRLS algorithm, it is preferable to have at least a few distinct variables in each set. The proposed methodologies for the estimation of (\ref{eqn:vc1}) and (\ref{eqn:vc2}) are also applicable to the COM-Poisson distribution's special cases (geometric for $\nu=0$, Poisson for $\nu=1$, and Bernoulli for $\nu \rightarrow \infty$), where the $\nu$ parameter needs to be fixed at the corresponding value in the IRLS algorithm.   


The tree-based algorithm approximates the varying coefficients $\bm{\beta}(\cdot)$, $\bm{\gamma}(\cdot)$ in Models (\ref{eqn:vc1}) and (\ref{eqn:vc2}) by a vectorial piece-wise constant function. Similar to \cite{burgin2015tree}, consider a partition of the value space $\mathcal{Z}_1 \times \cdots \times \mathcal{Z}_L $ of the $L$ splitting variables $Z_1, \ldots, Z_L$ into $M$ (terminal) nodes ${\mathcal{B}_1, \ldots,\mathcal{B}_M}$. The approximating true predictor functions at node $m$ are
\begin{eqnarray}
\begin{split}
\eta_{1i} &=&  1(\bm{z}_i \in \mathcal{B}_m) \left\{ \bm{x}_{1i}^{T}\bm{\beta}_m+ \bm{x}_{2i}^{T} \bm{\phi}_1  \right\},\\
\eta_{2i} &=& 1(\bm{z}_i \in \mathcal{B}_m) \left\{ \bm{w}_{1i}^{T}\bm{\gamma}_m + \bm{w}_{2i}^{T}\bm{\phi}_2 \right\},
\end{split}
\label{eqn:approx-pred}
\end{eqnarray} 
where $1(\cdot)$ is an indicator function that takes the value 1 if the argument is true and 0 otherwise. Based on (\ref{eqn:score-intro}), the estimated score functions for the varying coefficients of the predictor functions (\ref{eqn:approx-pred}) have the following form:
\begin{align}
  \begin{split}
\widehat{\bm{s}}_{1i} &=  1(\bm{z}_i \in \mathcal{B}_m)\bm{x}_{1i} (y_i- \widehat{E[y_i]}),\\
\widehat{\bm{s}}_{2i} &=  1(\bm{z}_i \in \mathcal{B}_m)\bm{w}_{1i} (-\ln(y_i!)+ \widehat{E[\ln(y_i!)]})\nu_i.
\end{split}
                          \label{eq:cmp-score}
\end{align}

The algorithm that we propose builds on the MOB framework by \citet{zei08}. Since the CMPMOB
formulation consists of  node invariant (global) parameters, we first obtain  consistent estimates for the global effects by fitting a global model
and then pass them along to all nodes of the fitted tree as an \emph{offset} term.  
This approach shares some similarities with the recent work of \citet{burgin2015tree} who developed  a tree-based varying coefficient regression for longitudinal data using a global random effect while the other fixed effects vary at each node. 

For the selection of splitting variable at each node, the MOB algorithm uses generalized M-fluctuation tests, also known as coefficient-constancy tests \citep{zei05}. The same tests can be applied to the estimated score functions in (\ref{eq:cmp-score}) to identify the splitting variable in the CMPMOB tree.
Based on the \emph{p}-values obtained from the test, the MOB algorithm identifies the
best splitting variable at each node after applying the Bonferroni or another 
correction for multiple testing. 

The next task is to estimate the optimal split point in the chosen
partitioning variable. The MOB algorithm uses an exhaustive search to do that;
it  fits two models for each potential split point candidate. Although
this is computationally intensive, there are efficient implementations available
for standard models (e.g., lmtree(), glmtree() functions in the R package \emph{partykit}
 \citep{hothorn1partykit}), but not for the  CMPMOB tree. To fill this void, we start by simplifying the exhaustive search
procedure and then  provide an alternative method to estimate the split point
by borrowing tools from the change point estimation framework. We discuss both 
approaches in greater detail in the following two subsections.

\subsection{Split Point Estimation Using Exhaustive Search}
Suppose the coefficient-constancy test is identified the splitting variable $z$
with  $n_z$ unique values. Then, the optimal split point 
is estimated based on
the following objective function:
\begin{eqnarray}
\underset{1 \le k \le n_z}{\text{argmin}} \left( -2 \ell(\widehat{\bm{\beta}}, \widehat{\bm{\gamma}}; \bm{y}, \bm{x}, \bm{z} \le \bm{z}_{(k)} )  -2 \ell(\widehat{\bm{\beta}}, \widehat{\bm{\gamma}}; \bm{y}, \bm{x}, \bm{z} > \bm{z}_{(k)} )\right),
\label{eqn:sp-exhaust}
\end{eqnarray}
where $\bm{z}_{(k)}$ is the $k$th order statistic for $\bm{z}$ and $\ell(\cdot)$ is the
log-likelihood function. 
For every $k$, two models need to be fitted to evaluate the two likelihood functions: $\ell(\widehat{\bm{\beta}}, \widehat{\bm{\gamma}}; \bm{y}, \bm{x}, \bm{z} \le \bm{z}_{(k)} )$ and $\ell(\widehat{\bm{\beta}}, \widehat{\bm{\gamma}}; \bm{y}, \bm{x}, \bm{z} > \bm{z}_{(k)})$, where $\bm{z}>\bm{z}_{(k)}:=(z_1>\bm{z}_{(k)},\ldots, z_n>\bm{z}_{(k)})^T$. This procedure is computationally heavy, especially for the CMPMOB tree. Since the  goal is to compare the  sum of $-2\ell$ values for different values of $k$, 
there is in fact no need to compute  the exact values;  approximate values serve the purpose. For example, for each model, the algorithm can run 
for a few fixed iterations and then the $-2\ell$ values are computed. 

The convergence of the 
IRLS algorithm for the CMP distribution is heavily dependent on the initial values chosen. Since a model is already fitted at each node, the estimated coefficients from that model can be used as the initial values for the models that are fitted for every potential split point in the exhaustive search (\ref{eqn:sp-exhaust}). The initial estimates are very close to the estimates obtained for each split point because it is the same data used for splitting and refitting the models. For this reason, only one or two iterations are required to evaluate models for each $k$, and this greatly reduces the computation time.

\citet{,wang2014boosted} discussed some heuristics to cut-down the number of unique values $n_z$ in either continuous or ordinal variables so that the number of model estimations would be reduced. Their approach is to specify a threshold $L$ (say 500 for instance) and then only consider the $L$ equally spaced quantiles as the unique split points if $n_z > L$. This approach seems reasonable and one could definitely try if the number of unique values is very large. 

In practice, it is possible to have 
some nominal partitioning variables with 
many categories, 
in which case exhaustive search can be computationally heavy. To simplify the exhaustive search in such cases, we transform the nominal variables into ordinal variables using the CRIMCOORD methods described in \citet{loh1997split}. Once the nominal variables are transformed to ordinal variables, the above heuristics can be used to reduce the number of  split points. However, care should be taken while implementing these heuristics for the CMPMOB tree as they may lead to loss of predictive power or loss of efficiency.

\subsection{Split Point Estimation Using Change Point Theory}
\label{sec:mob-method-cp}
For simplicity of exposition, 
we consider a linear predictor function $\eta_{1i}$ in Model (\ref{eqn:vc1}) and assume the other linear predictor $\eta_{2i}$ includes only an intercept. 
Suppose the coefficient-constancy tests on $\bm{\beta}$ identified variable $z$ as
the potential splitting variable, and that 
the true split point is
$\bm{z}_{(k)}$, the $k$th order statistic of $\bm{z}$. Now,  assume the \emph{true model} has the following form:
\begin{eqnarray}
  \eta_{1i} &=& 1(\bm{z} \le \bm{z}_{(k)}) \bm{x}_{1i}^{T}\bm{\beta}_L+ 1(\bm{z} > \bm{z}_{(k)}) \bm{x}_{1i}^{T}\bm{\beta}_R+\bm{x}_{2i}^{T} \bm{\phi}_1,
\label{eq:lp-true} 
\end{eqnarray}
where  $\bm{\beta}_L, \bm{\beta}_R$ are the
corresponding parameter vectors when $\bm{z} \le \bm{z}_{(k)}$ and $\bm{z} > \bm{z}_{(k)}$. 
In the first node  the \emph{working model} is 
of the form
\begin{eqnarray}
	\widetilde{\eta}_{1i} &=&  \bm{x}_{1i}^{T}\bm{\beta}+ \bm{x}_{2i}^{T} \bm{\phi}_1. 
	\label{eq:lp-working}
\end{eqnarray}
Then, from Equations 
(\ref{eq:lp-true}) and (\ref{eq:lp-working}) the \emph{true model} ($\eta_{1i}$) can be written as 
\begin{eqnarray*}
		\eta_{1i} &=&  \bm{x}_{1i}^{T}\bm{\beta}+\bm{x}_{2i}^{T} \bm{\phi}_1
                  +1(\bm{z} \le \bm{z}_{(k)}) \bm{x}_{1i}^{T}\bm{\beta}_L+1(\bm{z}
                  > \bm{z}_{(k)}) \bm{x}_{1i}^{T}\bm{\beta}_R - \bm{x}_{1i}^{T}\bm{\beta}\\
                  &=& \widetilde{\eta}_{1i} + \epsilon_i,
\end{eqnarray*}
where $\epsilon_i$, which is the omitted part from the \emph{working model}, is 
equal to
\begin{eqnarray*}
	\epsilon_i &=& 1(\bm{z} \le \bm{z}_{(k)}) \bm{x}_{1i}^{T}\bm{\beta}_L+1(\bm{z} > \bm{z}_{(k)}) \bm{x}_{1i}^{T}\bm{\beta}_R - \bm{x}_{1i}^{T}\bm{\beta}.
\end{eqnarray*}
In other words,
\begin{eqnarray}
	\epsilon_i &= & \bigg\{
	\begin{array}{@{}ll@{}}
	\bm{x}_{1i}^{T}\bm{\beta}_L-\bm{x}_{1i}^{T}\bm{\beta}, & \text{if}\
                                                                      \bm{z} \le
                                                                      \bm{z}_{(k)} \\
	\bm{x}_{1i}^{T}\bm{\beta}_R-\bm{x}_{1i}^{T}\bm{\beta}, & \text{if}\ \bm{z} > \bm{z}_{(k)}
	\end{array}, \quad i=1,\ldots,n.\label{eq:cp-eps}
\end{eqnarray}
In practice, $\epsilon_i$'s can be captured in the score functions that are calculated for the \emph{working model} (\ref{eq:lp-working}). From here onwards, we refer to the $i$th observation in each of the score functions as $\epsilon_i$. The formulation (\ref{eq:cp-eps}) makes it easy to deduce whether $\epsilon_i$  shows deviation
in either the mean or variance or both for values $\bm{z}> \bm{z}_{(k)}$. This
formulation resembles a change-point problem, which is extensively studied in the
literature \citep[e.g.][]{hin70,bou93,ant95,haw05,brodsky2013nonparametric}.

\citet{,ban07} clarify that change-point and split-point are not the same and the latter can be seen as complementary to the former. While 
change-point analysis assumes that there is a jump discontinuity in the underlying true model, it need not be the case for 
split-point estimation, and the underlying true function can be a smooth curve rather than 
having jump discontinuity. Hence, formally, change-point estimation methods might not be
useful for estimating split-points in the tree. Nonetheless, we can still
use some tools from change-point analysis to at least identify some potential splitting points, thereby allowing us to reduce the search space. This is very useful in practice.  

To test whether a change-point occurs at some location $k$, the estimated residuals $\widehat{\epsilon}_i, \quad i=1,\ldots,n,$ can be divided into two samples $\{ \widehat{\epsilon}_{z_{(1)}}, \ldots, \widehat{\epsilon}_{z_{(k)}}\}$ and $\{\widehat{\epsilon}_{z_{(k+1)}}, \ldots, \widehat{\epsilon}_{z_{(n)}}\}$, and we test whether both sets are identically distributed or not \citep{hin70}.
More formally, the null and alternative hypotheses can be written as
\begin{align}
H_0: \widehat{\epsilon}_{z_{(i)}} \sim F_0, \quad i=1,\ldots,n \\
H_1: \widehat{\epsilon}_{z_{(i)}} \sim \begin{cases}
F_0  \quad \text{if} \quad  i \le k, \\
F_1 \quad \text{if} \quad  i > k.
\end{cases}
\end{align}

A two-sample hypothesis test can be used to test for a change-point at ${z_{(k)}}$.  Let $D_{k,n}$ be the appropriately chosen test statistic, then the required split-point can be considered as $D_n = \underset{1 \le k \le n}{\text{max}} D_{k,n}$.  If we assume both $F_0$ and $F_1$ are Gaussian, then  $D_{k,n}$ is equivalent to the test statistic from the generalized likelihood ratio (GLR) test.

There are efficient implementations available for detecting change-points based on different assumptions (parametric or nonparametric). For example, the M-fluctuation framework proposed by \citet{zei05}  provides some tests that are similar to the change-point tests described above. However, those tests are very specific, model based, and do not provide the test statistic values for each potential split-point. Hence, for the sake of convenience, we consider the \emph{cpm} package \citep{ros13} in the  R software.  Since the proposed  procedure uses only  the score functions from the  estimated model at the parent node and there is no need to fit the models for each potential split-point, 
computational time can be reduced to a great extent.

Another advantage of the proposed change-point approach is that it can also be used to reduce
the number of potential candidates for the default exhaustive search. The top
5\% 
or 10\% 
of $D_{k,n}$ values can be considered instead of the exhaustive search, thereby 
significantly reducing the search space 
and simplifying computations. 
The complete CMPMOB estimation procedure is described in Algorithm \ref{alg:mob-cmp} in Appendix \ref{apx:algo}. 


Next, we evaluate 
the usefulness of the CMPMOB 
using an extensive simulation study.  

\subsection{Simulation Study 1: CMPMOB} 
\label{sec:mob-sim}
The purpose of this simulation study is to evaluate the usefulness of the CMPMOB semi-varying coefficient model and also to compare the performance of the proposed split point estimation procedure with the default exhaustive search method. For the sake of easy illustration, we present here the case where the same splitting variables are used for both models (\ref{eqn:vc1}) and (\ref{eqn:vc2}). 
The case where different splitting variables are used for each model is given in  Appendix \ref{sec:mob-ex2}.

The simulation design for this example is as follows:
\begin{itemize}
	\item Simulate $x_1, x_2, x_3, w_1,w_2 \sim U(0,1)$ and $z_1, z_2, z_3,z_4 \sim U(0,1)$. 
	\item Consider two smooth functions $f_1(x) = \text{sin}(2\pi x)$ and $f_2(x) = \text{cos}(2\pi x)$.
	\item Choose the split variable $z_1$ with split-point $0.65$ and compute the linear predictors $\eta_1=2+1(z_1>0.65)2x_1+1(z_1\le 0.65)x_2+2f_1^2(x_3)$ with $\lambda = \exp(\eta_1)$ and $\eta_2=0.25+1(z_1>0.65)0.5w_1+0.5f_2^2(w_2)$ with $\nu = \exp(\eta_2)$.
	\item Simulate $y \sim CMP(\lambda,\nu)$. 
\end{itemize}
 For  evaluation purposes,  four different sample sizes $n= \{500, 1000,
 2000, 5000\}$ are considered and  50 datasets are generated for each sample size. For each of these datasets, we fit the model  $\ln(\lambda) \sim
      \beta_0(z_1,z_2,z_3,z_4)+ \beta_1(z_1,z_2,z_3,z_4)x_1+\beta_2(z_1,z_2,z_3,z_4)x_2+
      s(x_3); \ln(\nu) \sim  \gamma_0(z_1,z_2,z_3,z_4)+ \gamma_1(z_1,z_2,z_3,z_4)w_1+
      s(w_2) $  with $z_1$, $z_2$, $z_3$ and $z_4$ as the
 potential moderator/splitting variables and $s(x_3)$, $s(w_2)$ as fixed effects/global terms.  
 %
%
 %
 Split-points are estimated via:
 (1) exhaustive search, (2) search through a set of exact change-points for each score function, and (3) search through a set of the top 10\% change-points for each score function. For each  model, the estimated split-point, the sum of local $-2\ell$ values of the fitted models at all terminal nodes, and the  number of terminal nodes are recorded. 
 
\begin{figure}[h]
 	\begin{minipage}{0.5\textwidth}
 		\includegraphics[width=1\linewidth]{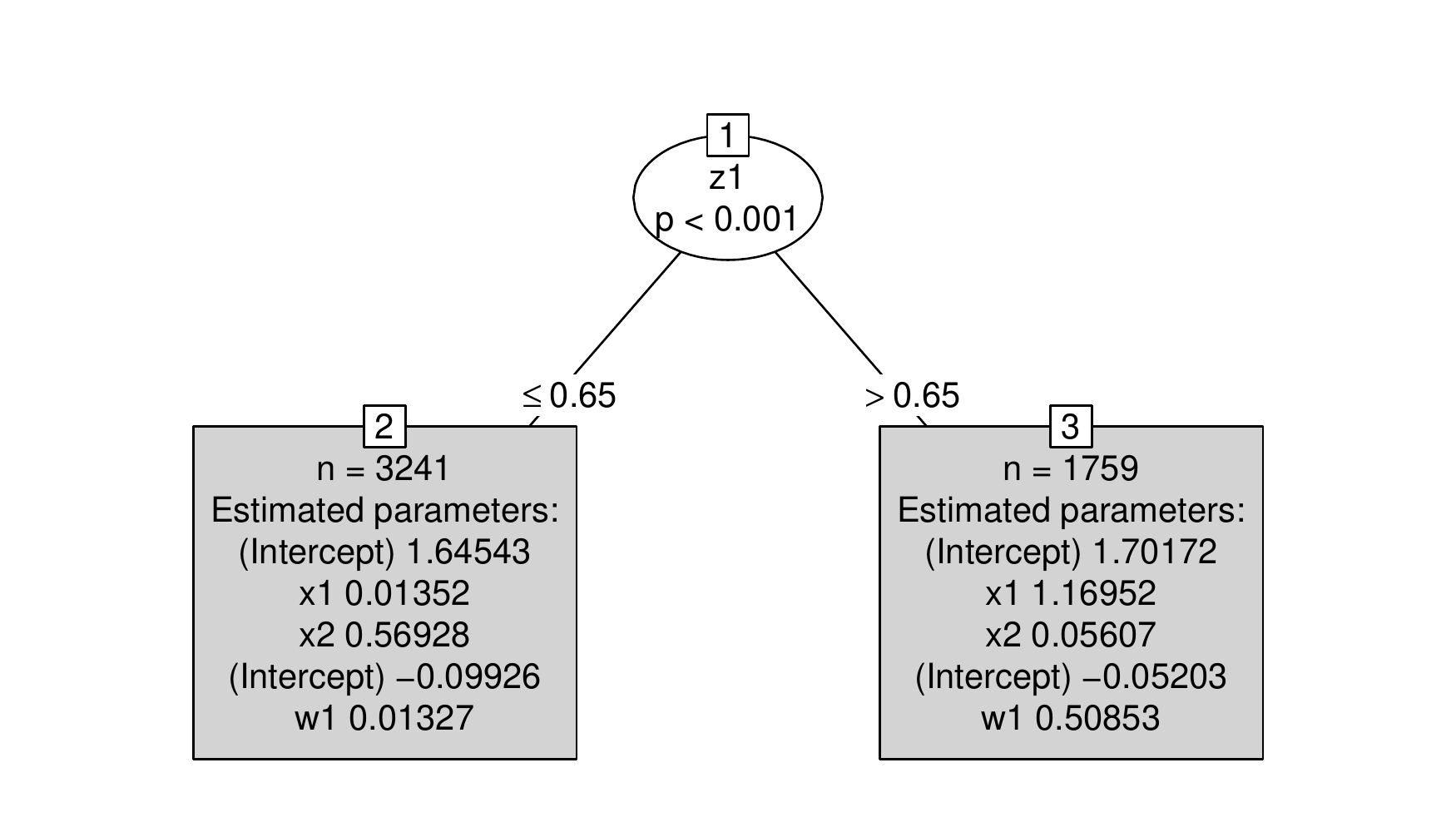}
 	\end{minipage}
 	\begin{minipage}{0.5\textwidth}
 		\includegraphics[width=1\linewidth]{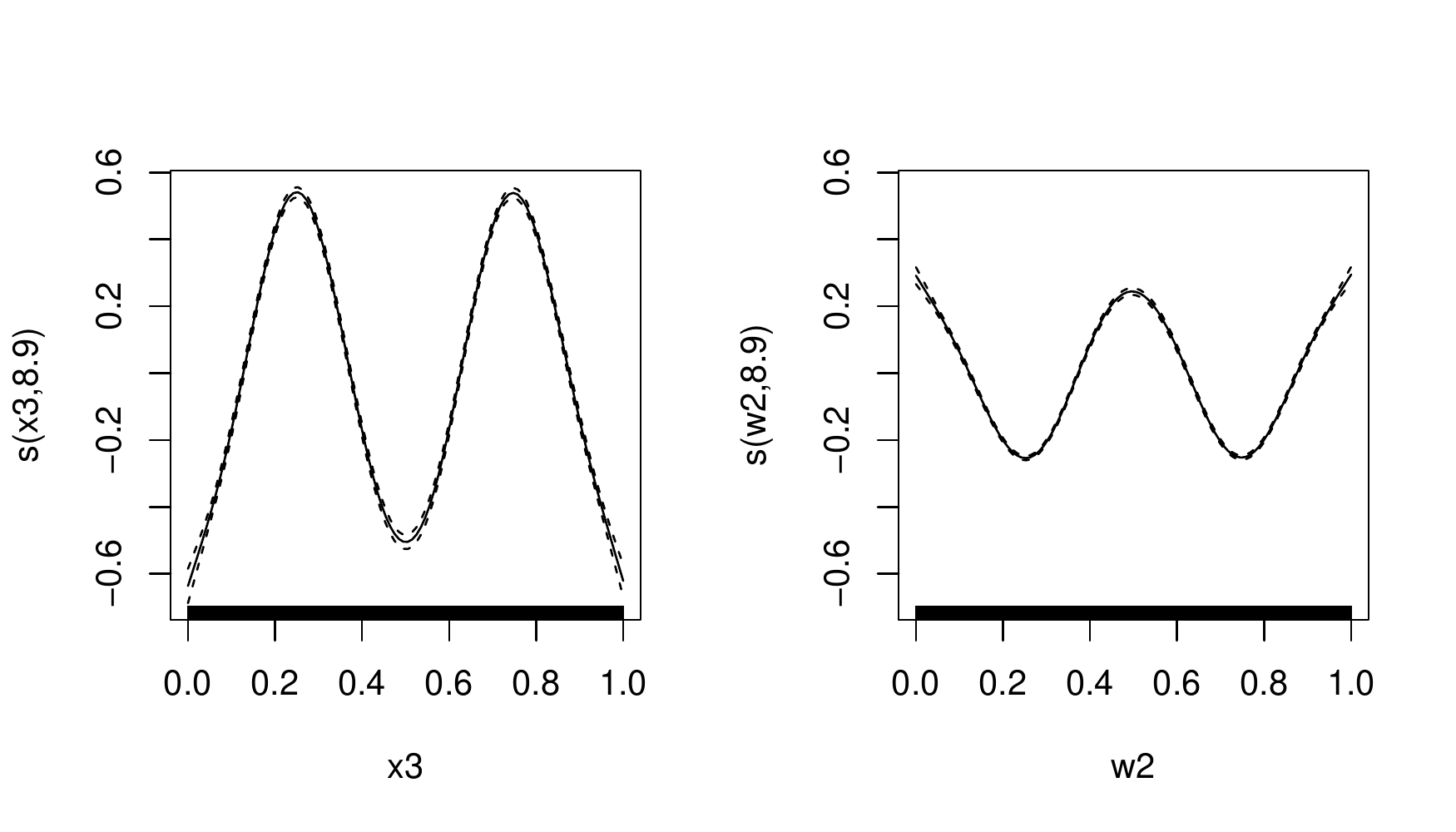}
 	\end{minipage}
 	\caption{Left: Fitted MOB tree. 
 	Right:  fixed effects $2\text{sin}^2(2\pi x_3)$ and $0.5\text{cos}^2(2\pi w_2)$.}
 	\label{fig:mob-tree}
 \end{figure}

 The estimated model for one of the simulated datasets with $n=5000$ is shown in Figure \ref{fig:mob-tree}. The results include the smooth terms for both $\ln(\lambda)$ and $\ln(\nu)$ models and the tree.
 
\begin{figure}
 \centering
\includegraphics[width=1\textwidth]{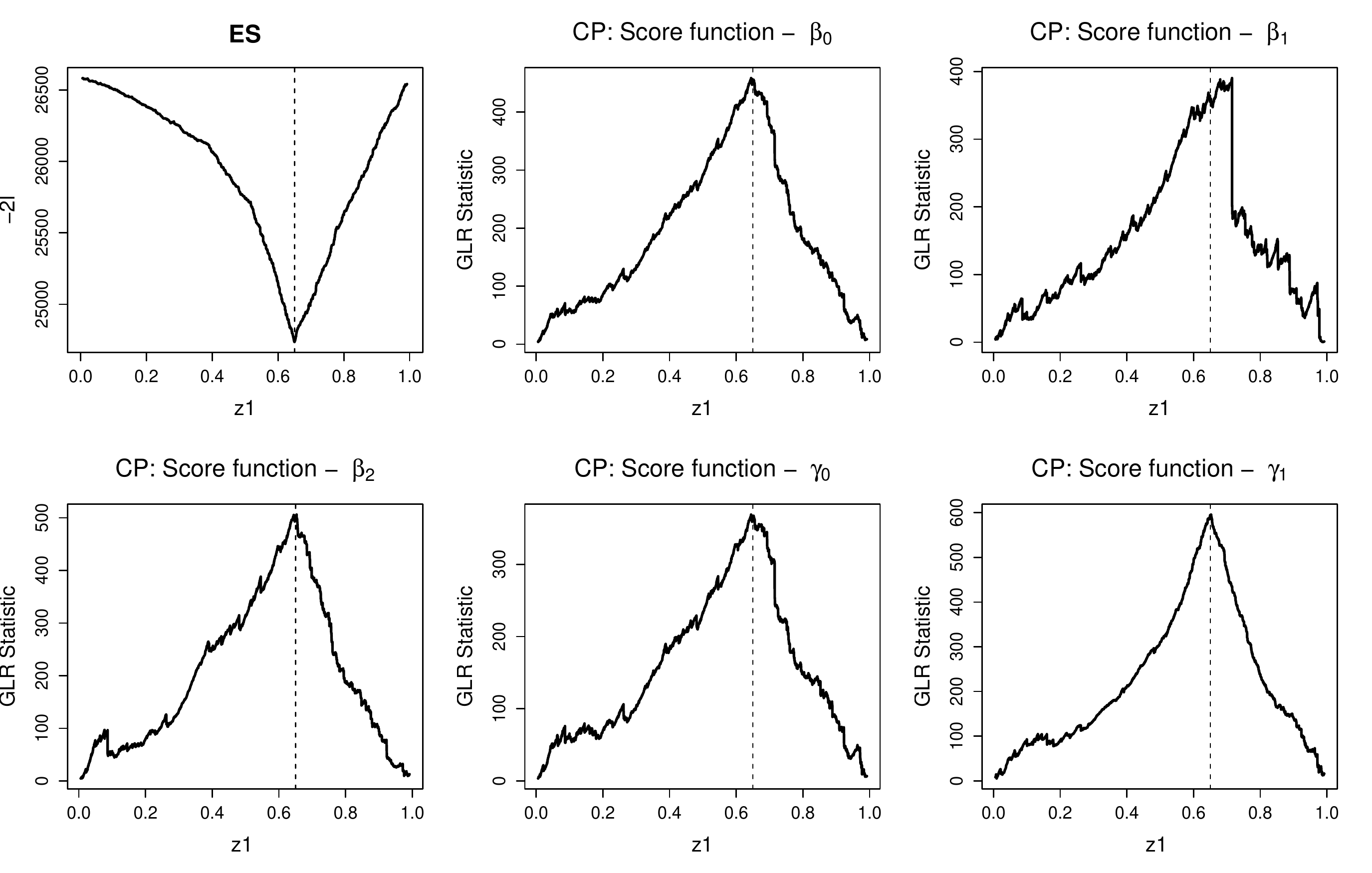}
\caption{Estimated $-2\ell$ values for all the models and the estimated GLR test statistics for each score function, evaluated at each potential split-point in the chosen moderator variable $z_1$. The dashed vertical line is the true split-point located at $z_1=0.65$.}
\label{fig:ES-2l}
\end{figure}

The usefulness of the change-point estimation procedure 
is shown in Figure \ref{fig:ES-2l}. When the data contains a clearly defined split-point, as in this case, it is identified by the maximum or one of the 
change-point statistics  (in this case GLR test) calculated for each score function. It is evident from  Figure \ref{fig:ES-2l} that the true split-point $z_1=0.65$ is identified by the GLR test statistics for each score function except for the score function for $\beta_1$. Even in that case, the top 5\% or 10\% of the test statistics contain the true split-point.

\begin{figure}
	\centering
	\includegraphics[scale=0.5]{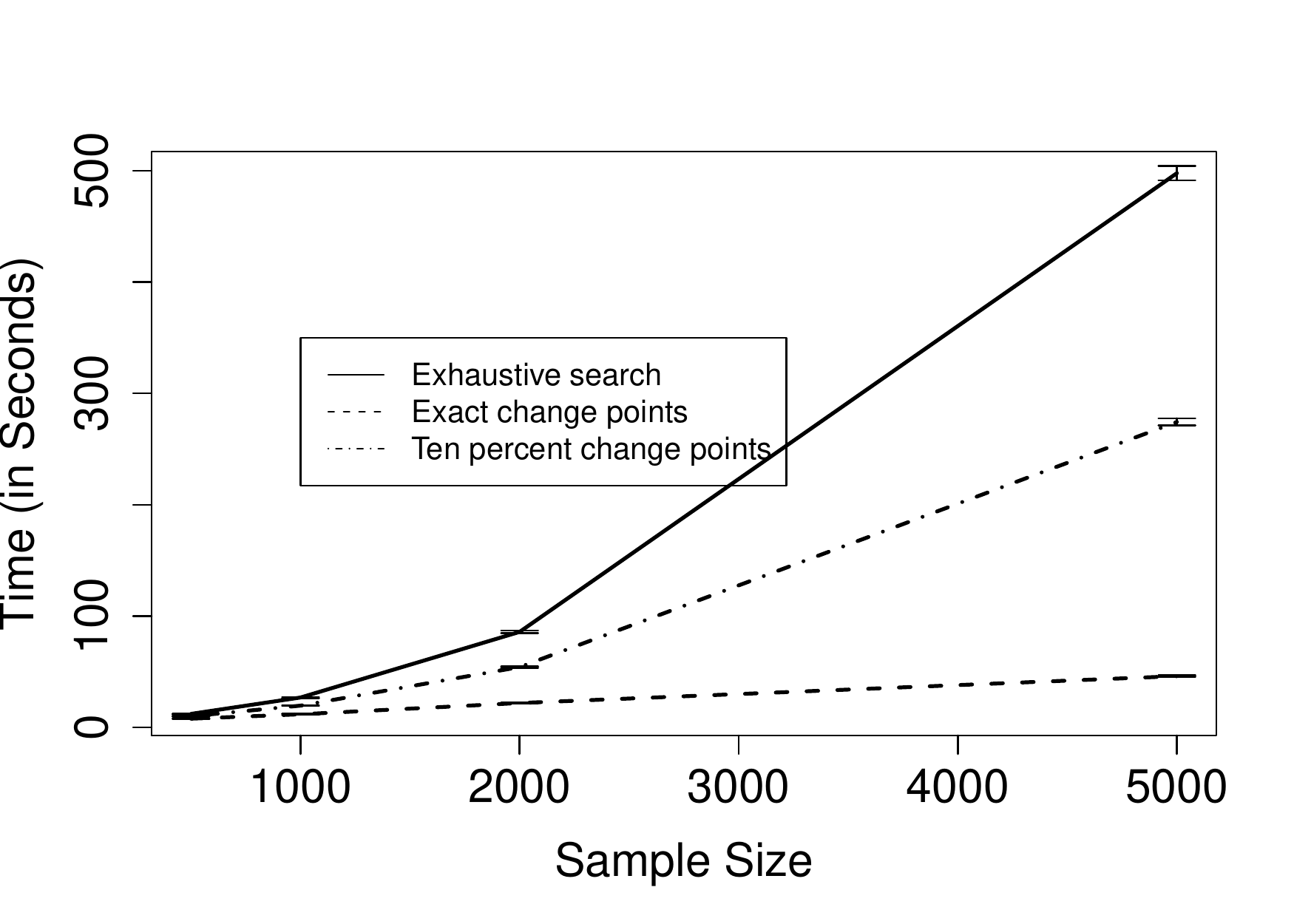}
	\caption{Computation times for the CMPMOB tree estimated using  exhaustive search, the change-point method with exact set of change-points, and top 10\%. Standard errors are calculated based on 50 simulations. Computed on a Linux machine with 32GB RAM.}
	\label{fig:mob-comput}
\end{figure}

 The full simulation results are described in Table
 \ref{tab:mob-sim-wglo}. 
 The results using
 the exhaustive search and the search with 10\% change-points are identical 
 except for one dataset with sample size $n=500$. 
 We find that the cut-off of 10\% is mostly unnecessary 
 as the cut-off of 5\% suffices. 
 While the results are slightly different for the models estimated with the exact set of change-points and the set of 10\% change-points, in terms of the overall fit measures (local $-2l$), they are nearly identical. The computational times for the three models  across all the simulated datasets are shown in Figure \ref{fig:mob-comput}. Not surprisingly, the models based on change-point estimation methods are much faster compared to those based on exhaustive search, although the latter were made faster by using only one iteration from the
 IRLS for each potential split-point.

  \begin{table}[!htbp]
  \caption{\label{tab:mob-sim-wglo} Model  $\ln(\lambda) \sim
      \beta_0(z_1,z_2,z_3,z_4)+ \beta_1(z_1,z_2,z_3,z_4)x_1+\beta_2(z_1,z_2,z_3,z_4)x_2+
      s(x_3); \ln(\nu) \sim  \gamma_0(z_1,z_2,z_3,z_4)+ \gamma_1(z_1,z_2,z_3,z_4)w_1+
      s(w_2) $  with the smooth $s(x_3), s(w_2)$ as fixed/global effects. The numbers in parentheses are standard deviations across 50 simulations. } 
  \small
  	\begin{tabular}{ccccc}
  		\hline
  		
  		& \multicolumn{4}{c}{Exhaustive Search} \\
  		\cline{2-5}  
  		& $n=500$ & $n=1000$ & $n=2000$ &$n=5000$ \\
  		\hline 
  		Split-1& $0.646(0.009)$  & $ 0.649(0.003)$ & $0.650(0.002)$ & $0.650(0.0007)$  \\ 
  		Global $-2l$ & $2499.08(37.32)$   & $5017.50 (58.33)$  & $10036.51 (84.45)$ & $25156.85 (141.98)$  \\ 
  		Local $-2l$ & $2245.23 (27.49)$   & $4481.65 (54.70)$  & $ 8964.27 (72.68)$ & $22434.99 (118.67)$  \\
  		No.of terminal nodes & $2$ & $2$ & $2 $ & $2$ \\
  		\hline 
  		
  	\end{tabular} 
  	\begin{tabular}{ccccc}
  		\hline
  		&  \multicolumn{4}{c}{Search with Exact Change-Points} \\
  		\cline{2-5}  
  		& $n=500$ & $n=1000$ & $n=2000$ &$n=5000$  \\
  		\hline 
  		Split-1&  $0.653(0.016)$  & $ 0.651(0.005)$ & $0.650(0.003)$ & $0.650(0.001)$  \\ 
  		Global $-2l$ & $2499.33(38.37)$   & $5016.60 (58.65)$  & $10036.51 (84.45)$ & $25156.85 (141.98)$  \\ 
  		Local $-2l$ & $2247.66 (28.71)$   & $4486.22 (56.50)$  & $8966.11 (71.83)$  & $ 22437.75 (117.95)$  \\
  		No.of terminal nodes & $2$ & $2$ & $2 $ & $2$ \\
  		\hline 
  		
  	\end{tabular}
  	\begin{tabular}{ccccc}
  		\hline
  		&  \multicolumn{4}{c}{Search with 10\% Change-Points} \\
  		\cline{2-5}  
  		& $n=500$ & $n=1000$ & $n=2000$ &$n=5000$  \\
  		\hline 
  		Split-1&  $0.646(0.01)$  & $ 0.649(0.003)$ & $0.649(0.0012)$ & $0.650(0.0007)$  \\ 
  		Global $-2l$ & $2499.07(37.32)$   & $5017.50 (58.33)$  & $10036.51 (84.45)$ & $25156.85 (141.98)$  \\ 
  		Local $-2l$ & $2245.54 (27.48)$   & $4481.65 (54.70)$  & $8964.27 (72.68)$  & $22434.99 (118.67)$  \\
  		No.of terminal nodes & $2$ & $2$ & $2 $ & $2$ \\
  		\hline 
  		
  	\end{tabular}

  \end{table}
 The  results from Table \ref{tab:mob-sim-wglo} 
 and Figure \ref{fig:mob-comput} 
 indicate that when there exists a clear split-point in the data, as in the case of the simulated datasets, change-point estimation methods are useful for 
 reducing the computational complexity. 
 Moreover, 
 differences from split-points 
 estimated with the exact set of change-points 
 do not necessarily imply estimation error, and in some cases 
 those split-points make better sense practically, when 
 interpreting the results. In practice, we suggest 
 to first fit the model using the exact set of change-points and only then 
 proceeding with the top 10\% change-points or the exhaustive search method.


Appendix  \ref{sec:mob-ex2} provides a similar simulation, but where the moderator variables for $\ln(\lambda)$  and $\ln(\nu)$ models are different.
 The results and conclusions 
 are similar to those from the above example with the same moderators. When the moderator variables for $\ln(\lambda)$  and $\ln(\nu)$ models were different, 
 we noticed that the CMPMOB 
 generated 3 to 4 partitions to accommodate those different splits. This is one of the limitations of CMPMOB,  
 which is due to the dependence of the $\ln(\lambda)$  and $\ln(\nu)$ models. More specifically, both models are estimated using the same data at each terminal node and they are 
 not expected 
 to have different estimated splits. In the next section, we propose a generalized model that overcomes this limitation and provides smoother approximations to the coefficient functions.

\section{CMPBoost Semi-Varying Coefficient Model}
\label{sec:gb}

\subsection{Background}
The CMPMOB 
model usually produces discontinuous and piece-wise
approximations to the coefficient functions $\bm{\beta}(\bm{z}_i)$ or $\bm{\gamma}(\bm{z}_i)$. 
Gradient boosting \citep{,freund1996experiments,freund1997decision,friedman2001greedy} is a popular method that can provide  more smoother estimates than a single tree. Gradient boosting is a recursive nonparametric
algorithm with successful applications in many areas. The key idea is to
combine a large number of base learners (such as simple trees) to achieve
better predictive accuracy. For a comprehensive overview, 
see \citet{,friedman2001greedy} and \citet{buehlmann2006boosting}.

Recently, \citet{,wang2014boosted} proposed a gradient boosting approach for the tree-based varying coefficient model. Our approach follows along the same 
lines but is more involved due to the complicated nature of the CMP
distribution. 
We first briefly describe the approach
proposed by \citet{,wang2014boosted} 
and then propose our approach for the CMPBoost semi-varying coefficient model.

Consider the varying coefficient model:
$y_i=f(\bm{x}_i,\bm{z}_i)+\epsilon_i = \bm{x}_i^T\bm{\beta}(\bm{z}_i)+\epsilon_i$.
By choosing an empirical loss function $\phi(y,f)$ such as least square error or
absolute error loss, the model $f$ can be estimated as follows:
\begin{align}
  \widehat{f} &= \underset{f \in \mathcal{F}}{\text{arg min}}\frac{1}{n}\sum_{i=1}^n \phi(y_i,f(\bm{x}_i,\bm{z}_i)),
  \end{align}
  where $\mathcal{F}=\{f(\bm{x},\bm{z})\vert
  f(\bm{x},\bm{z})=\bm{x}^T\bm{\beta}(\bm{z})\}$ is the constrained function
  space. 
  In boosting, the goal is to find an incremental model $T(\bm{x}_i,\bm{z}_i)$ 
  that minimizes empirical risk,
  \begin{align*}
    \widehat{T} &= \underset{f \in \mathcal{F}}{\text{arg min}}\frac{1}{n}\sum_{i=1}^n \phi\left(y_i,\widehat{f}^{[b-1]}(\bm{x}_i,\bm{z}_i)+T(\bm{x}_i,\bm{z}_i)\right),
  \end{align*}
where $\widehat{f}^{[b-1]}$ is the model fit from the $[b-1]$th iteration.
  The gradient boosting algorithm starts with a simple fit $\widehat{f}^{(0)}$
  and then iteratively updates the estimate by adding the incremental model
  fitted on ``residuals'',  obtained by evaluating the negative gradient
  at the current fitted tree $u_i=-\frac{\partial}{\partial
    f}\phi(y_i,f)\vert_{f=\widehat{f}}$. The final boosting model
  has the following form:
  \begin{align}
    \widehat{f}^{(B)} &= \widehat{f}^{(0)}+\xi \sum_{b=1}^B\sum_{m=1}^M\bm{x}_i^T\widehat{\bm{\beta}}_m^{(b)}\bm{1}_{(\bm{z}_i\in \widehat{C}_m^{(b)})},
                        \label{eqn:b-update}
  \end{align}
  with $\widehat{C}_m^{(b)}$ denoting the terminal node for the tree produced in the
  $b$th boosting iteration and $B$ denotes the number of boosting steps. According to  \citet{,wang2014boosted}, the boosting algorithm involves four
  tuning parameters: the number of boosting iterations $B$, the learning rate
  $\xi$, the minimal node size, and the size of each base learner $M$. Based on
  \citet{,friedman2001greedy}, one can use small values for $0 < \xi <1$, for example
  $\xi=0.005$, to achieve better predictive accuracy.

  The gradient boosting algorithm can be generalized to nonlinear and
  generalized linear models using different loss functions such as absolute  error loss, Huber-loss \citep{,buhlmann2007boosting}, or likelihood. In
   standard generalized linear models, the mean is related to the covariate
  function through a link function. Owing to this, one could follow the same
  approach as above to implement the gradient boosting algorithm when there is only one parameter. 
  However, when there is a nuisance parameter in the likelihood, one possible option is
  to use the gradient boosting algorithm for the main parameter and later estimate
  the nuisance parameter directly. For example,  \citet{,yang2017insurance} provided the gradient boosting algorithm for
  Tweedie compound Poisson models. The authors first fix the nuisance parameter
  and estimate the mean parameter through gradient boosting. Once the
  mean parameter is estimated, they
  use the profile likelihood approach to estimate the nuisance parameter. While
  this approach works for the Tweedie compound Poisson distribution, it might not be appropriate
 for the CMP distribution because it is not possible to fix the $\nu$ parameter and estimate the $\lambda$ parameter (or $\mu$ parameter in \citet{,huang2017mean}) separately. 
  For this reason, we consider updating both parameters in the  gradient boosting algorithm. 

  \subsection{CMPBoost Estimation}
 We propose an estimation procedure for the CMPBoost 
 model using 
 regression trees as  base learners. 
 Similar to the IRLS 
 algorithm for estimating CMP regression (see Section \ref{sec:back}), 
 our proposed approach updates both the
 parameters $\lambda$ and $\nu$. 
 For this two-step estimation, we consider the following model formulation that is even more flexible than the one in (\ref{eqn:vc1}, \ref{eqn:vc2})
 \begin{eqnarray}
\label{eqn:bvc1}
\ln(\lambda_i)= \eta_{1i} = \bm{x}_{1i}^{T}\bm{\beta}(\bm{z}_{i})+ \bm{x}_{2i}^{T} \bm{\phi}_1, \\
\ln(\nu_i)= \eta_{2i} = \bm{w}_{1i}^{T}\bm{\gamma}(\bm{u}_{i})+ \bm{w}_{2i}^{T}\bm{\phi}_2,
\label{eqn:bvc2}
\end{eqnarray} 
 where $\bm{u}_i=(u_{i1},\ldots,u_{iM})^T$ is another set of moderator variables. In practice, the moderator variables in both the models,  $\bm{z}_i$ and $\bm{u}_i$, can be the same or completely different. Note that the model formulation for the CMPMOB tree uses the same set of moderators for both (\ref{eqn:vc1}) and (\ref{eqn:vc2}) as both parameters need to be estimated at each terminal node using the same data. For this reason, CMPMOB tree is limited in the sense that it does not support formulation (\ref{eqn:bvc1}) and (\ref{eqn:bvc2}). On the other hand, the CMPBoost 
 model is flexible enough to allow different moderators for each of (\ref{eqn:bvc1}) and (\ref{eqn:bvc2}), although 
 model interpretation 
 becomes 
 more complicated.

 Using the notation from Section \ref{sec:back}, the  adjusted dependent
 variable for the $\ln (\lambda_i)$ model in (\ref{eqn:bvc1}) for the $b$th iteration is expressed as
 \begin{align*}
\widehat{t}_{i1}^{(b)}=\widehat{\eta}_{i1}^{(b-1)}+\frac{y_i-\widehat{E[y_i]}^{(b-1)}}{\widehat{V[y_i]}^{(b-1)}},
 \end{align*}
 where $\widehat{E[y_i]}^{(b-1)}$ and $\widehat{V[y_i]}^{(b-1)}$ are the
 estimated  mean and variance for the $b$th iteration. The linear predictor is
 $\widehat{\eta}_{i1}^{(0)}=\widehat{f}_{i1}^{(0)}=\bm{x}_{1i}^{T}\widehat{\bm{\beta}}^{(0)}+\bm{x}_{2i}^{T}\widehat{\bm{\phi}}_1$.
As in the gradient boosting approach, the linear predictor $\widehat{\eta}_{i1}^{(0)}$ is updated using the
base learners fitted with the estimated residual
$\frac{y_i-\widehat{E[y_i]}^{(b-1)}}{\widehat{V[y_i]}^{(b-1)}}$ as the dependent
variable. 

Similarly, the adjusted dependent variable for the $\ln(\nu_i)$ model in (\ref{eqn:bvc2})
for the $b$th iteration is expressed as:
\begin{align*}
  \widehat{t}_{i2}^{(b)}=\widehat{\nu}_i^{(b-1)} \widehat{\eta}_{i2}^{(b-1)}+\frac{-\ln(y_i!)+\widehat{E[\ln(y_i!)]}^{(b-1)}}{\widehat{V[\ln(y_i!)]}^{(b-1)}},
  \end{align*}
  where $\widehat{E[\ln(y_i!)]}^{(b-1)}$ and $\widehat{V[\ln(y_i!)]}^{(b-1)}$ are the estimated mean and variance for  $\ln(y_i!)$ in the $b$th iteration.
If the $\ln(\nu_i)$ model also contains varying coefficients, the same approach is used 
to update the linear predictor $\widehat{\nu}_i^{(0)} \widehat{\eta}_{i2}^{(0)}=\widehat{f}_{i2}^{(0)}$. The base learners are fitted to the estimated residual
$\frac{-\ln(y_i!)+\widehat{E[\ln(y_i!)]}^{(b-1)}}{\widehat{V[\ln(y_i!)]}^{(b-1)}}$
for updating the linear predictor similar to  (\ref{eqn:b-update}).  

Since the estimated residual is a continuous variable, any regression tree algorithm can be used to fit the base learner.   We consider the ``PartReg" algorithm \citep{,wang2014boosted} as it is easy to control its tuning parameters and simpler to evaluate its model predictions. For example, it is not 
straightforward to apriori limit the number of terminal nodes in a MOB tree, since the 
algorithm first grows a larger-than-necessary tree 
and then prunes it back to the desired size. In contrast, 
limiting the number of terminal nodes is 
easy with the PartReg algorithm as it uses breadth-first search cycles. 
The PartReg algorithm does have some drawbacks, such as its 
selection bias towards variables with many split points \citep{strobl2008conditional} and 
computational challenges 
for identifying split points in  ordinal and nominal variables (see \citet{wang2014boosted} for a detailed discussion). Nonetheless, we still consider the PartReg algorithm as it is convenient, and we leave the MOB algorithm implementation of the CMPBoost 
model for future work.

 
The boosting algorithm has four tuning parameters, namely, $\xi$, $M$, $B$ and the minimal node size. Among these four, we fix $\xi=0.01$,  $0.05$, $0.1$ or $0.5$, and the minimal node size as $20$ in our simulation studies and real examples. The number of boosting iterations $B$ depends on the value chosen for $\xi$; a smaller value leads to more 
iterations. 
The choice of both $\xi$ and $B$ is data-dependent.\footnote{Care should be taken while setting the value for $\xi$, as large values can sometimes lead to non-convergence 
of the algorithm. It is therefore advisable to first experiment with a few values and observe the convergence path (sequence of $-2l$ values) and number of iterations.} 
We use the $-2$\emph{log likelihood}
 evaluation criterion and stop the algorithm when there is not much improvement in the evaluation criterion, and hence $B$ is determined automatically. If a very small value is chosen for the parameter $\xi$, we usually set $B$ to a fixed number, say $1000$.  Finally, the parameter $M$, the number of terminal nodes, is evaluated based on the performance on the test data.

 As  mentioned, one of the major limitations of the CMPMOB
 model is its computational complexity. However, the proposed CMPBoost model  is computationally  simpler than the CMPMOB, as the
 base learners are regular regression trees which are 
 faster to compute.
  To interpret the results from  CMPBoost, 
  one can use variable
 importance plots and  partial dependence plots as suggested by 
 \citet{,wang2014boosted}. Following \citet{,friedman2001greedy} and \citet{,hastie2015statistical}, the
 importance of variable $z_j$ in a single tree $(T)$ is defined as 
 \begin{align*}
   \mathcal{I}_j^2(T) &= \sum_{l=1}^{M-1} \Delta SSE_l \bm{1}_{(v(l)=z_j)},
 \end{align*}
 where $v(l)$ denotes the variable chosen for splitting in the $l$th step, $M$ is
 the number of terminal nodes, and
 $\Delta SSE_l$ is the improvement in the $l$th iteration of the model fit. In boosting, we denote the $b$th tree as $T_b$ and compute the relative importance of $z_j$ as the average of $\mathcal{I}_j^2(T_b)$ among all iterations, namely,
 \begin{align*}
   \mathcal{I}_j^2 &= \frac{1}{B}\sum_{b=1}^B \mathcal{I}_j^2(T_b).
 \end{align*}
 Coefficient interpretation has always been  
 tricky 
 with CMP models. Furthermore, it is also not 
 easy to construct partial dependence plots for the proposed CMPBoost 
 due to the two models, $\ln(\lambda_i)$, $\ln(\nu_i)$, which are not orthogonal and their updates depend on each other. We illustrate some of these issues in a 
 simulation study in the following section.  

 \subsection{Simulation Study 2: CMPBoost}
 \label{sec:gb-sim}

 The goal of this simulation study is to showcase the flexibility of 
CMPBoost. 
We examine a case that
includes varying coefficients for both intercept and slope terms in the $\ln(\nu)$ and $\ln(\lambda)$ models with different moderator variables. 


  The simulation design is as follows:
  \begin{itemize}
\item Generate $x_i, w_i \sim U(0,1)$ and $\bm{z}_i=(z_{i1},\ldots,z_{i,10})^T $, where  $z_{ij} \sim U(0,1)$, $j=1,\ldots,10$.
\item Compute the intercept functions $\beta_0(\bm{z}_i)=\text{sin}^2(2\pi
  z_{i1})+\text{exp}(z_{i2}-1)$, $\gamma_0(\bm{z}_i)=\text{sin}^2(2\pi
  z_{i5})$ and the slope functions
  $\beta_1(\bm{z}_i)=2\text{cos}^2(2 \pi z_{i1})+z_{i2}(1-z_{i2})$, $\gamma_1(\bm{z}_i)=0.5\text{cos}^2(2 \pi z_{i6})$.
\item Generate the dependent variable  $y_i \sim CMP(\lambda_i=\text{exp}\left(
    \beta_0(\bm{z}_i)+x_i\beta_1(\bm{z}_i) \right),
  \nu_i=\text{exp}\left(
    \gamma_0(\bm{z}_i)+w_i\gamma_1(\bm{z}_i) \right))$.
\end{itemize}
 To allow sparsity, we let the true functions $\beta_0(\bm{z}_i)$ and $\beta_1(\bm{z}_i)$ only depend on the moderator variables $z_1$ and $z_2$. Similarly, the true functions $\gamma_0(\bm{z}_i)$ and $\gamma_1(\bm{z}_i)$  depend on $z_5$ and $z_6$, respectively.  The total  sample size   is $n=1000$, and we leave 40\% of the observations for the test data while the remaining 60\% is used as training data. The tuning parameter $M$, the number of terminal nodes for each  base learner,  is determined based on the out-of-sample prediction error (or $-2l$).  In general,  the algorithm iterates until there is not much change in the $-2l$ value or for a predefined number of iterations,  $500$, and hence $B$ is determined automatically.  The minimum node size is chosen as $20$ and the learning parameter is set as $\xi=0.1$. 

\begin{figure}
    \centering
    \vspace{-0.5in}
    \includegraphics[width=0.5\textwidth]{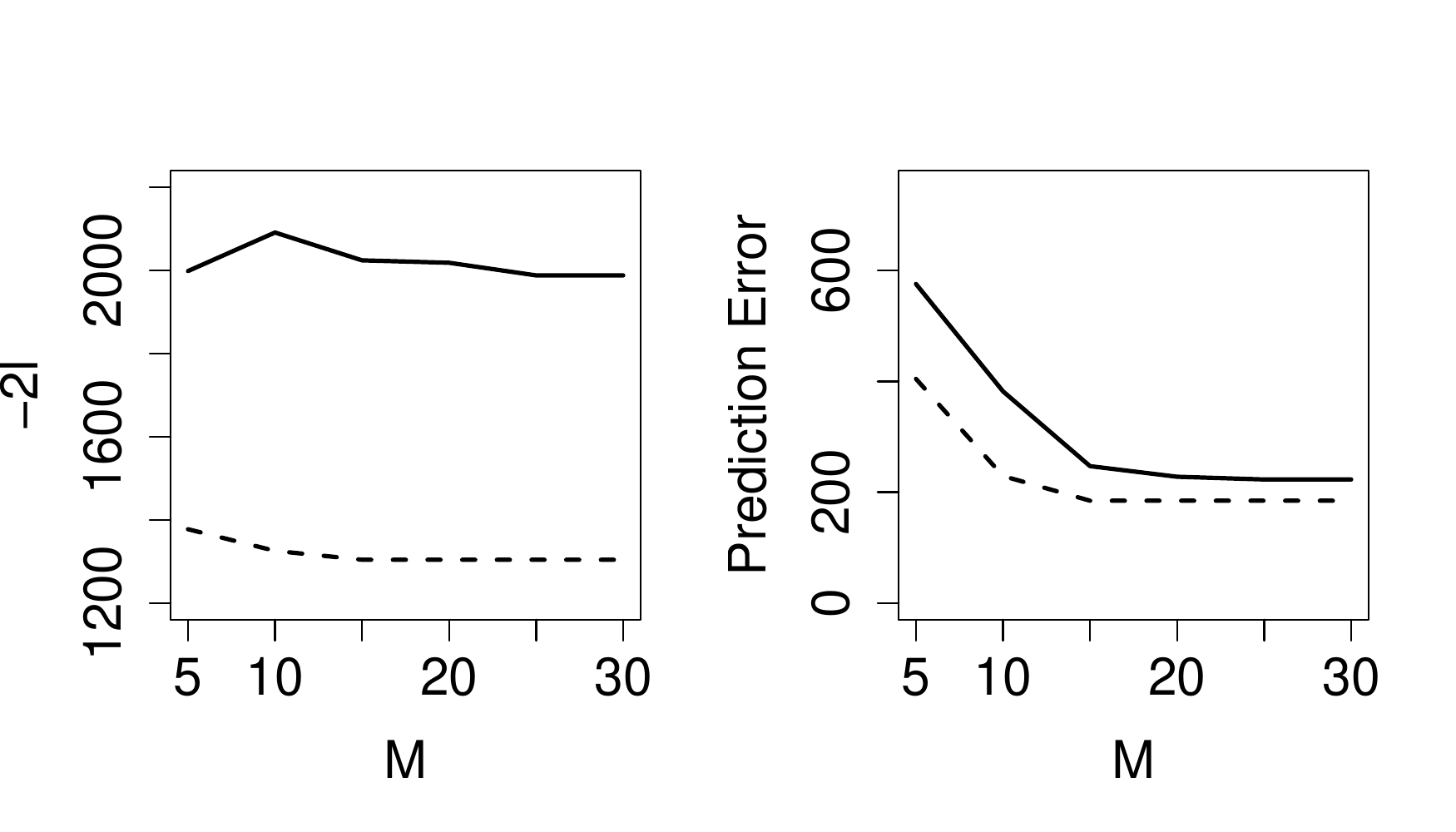}
    \caption{\textbf{Left:} -2 log-likelihood values. \textbf{Right:} prediction error for the models with an increasing number of terminal nodes ($M$). Solid line represents the values for the training data and dashed line for the test data. In both plots, the minimum for test data is obtained at $M=15$ and  $\xi=0.1$ for each model which took $170$ iterations, on an average, to converge.}
    \label{fig:gb-train-m-ex2}
\end{figure}

Figure \ref{fig:gb-train-m-ex2} shows the results for the models fitted with an increasing number of terminal nodes for base learners ($M$) on both training and test data. 
Both $-2l$ and prediction error have similar behavior and obtain their minimum at $M=15$ on the test data.\footnote{To remove possible overfitting associated with the number of boosting iterations, for each $M$ we let the model fitted to the training data run until there is not much change in $-2l$, and we fix the same number of iterations for the model on the test data for that specific $M$. For example, if the model with $M=15$ takes $170$ iteration for the training data, the same $170$ iterations are used to evaluate the prediction error on the test data.}
 We simulated 20 data sets with sample size $n=1000$ and constructed the partial dependence plots. Interpreting the coefficient functions from the CMP model is not straightforward. As discussed in Section \ref{sec:back}, the coefficients for the $\ln \lambda$ model depends on the $\nu$ parameter and  are approximately scaled down by $\nu$ in terms of the mean parameter. Nevertheless, we constructed the partial dependence plots, for each varying coefficient function, analogous to the approach used by \citet{,wang2014boosted}. 

\begin{figure}[!htbp]
\centering
\begin{minipage}{1\textwidth}
\centering
  \includegraphics[width=0.95\textwidth]{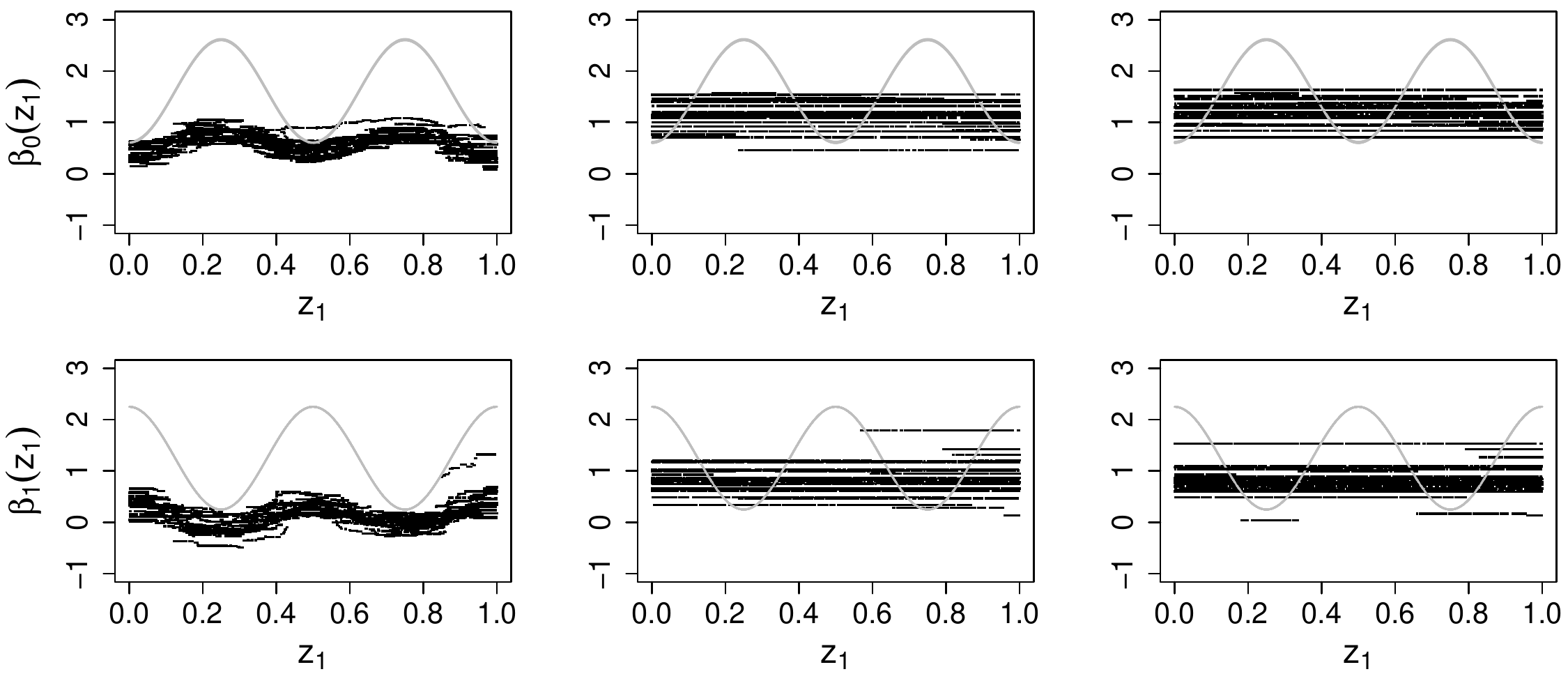}
\caption{Reconstructed varying coefficient surfaces for $\beta_0(\bm{z})$ (top row) and  $\beta_1(\bm{z})$ (bottom row) for boosting, MOB tree fitted with exhaustive search,  and MOB tree fitted with change point (10\% points), using 20 simulations.}
\label{fig:boost-fn1-sim2}
\end{minipage}
\begin{minipage}{1\textwidth}
\centering
  \includegraphics[width=0.95\textwidth]{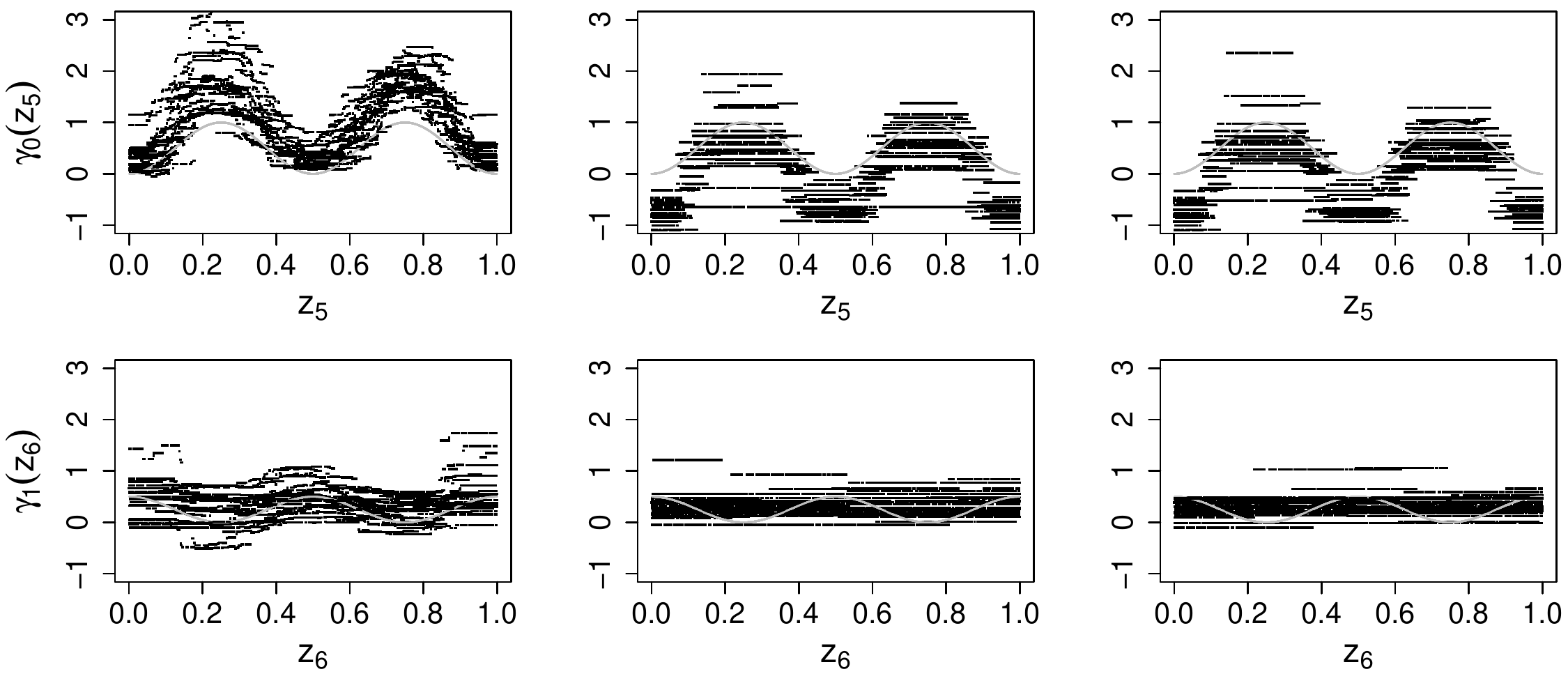}
\caption{Reconstructed varying coefficient surfaces for $\gamma_0(\bm{z})$ (top row) and  $\gamma_1(\bm{z})$ (bottom row) for boosting, MOB tree fitted with exhaustive search, and MOB tree fitted with change point (10\% points), using 20 simulations.}
\label{fig:boost-fn2-sim2}
\end{minipage}
\end{figure}

Figure
\ref{fig:boost-fn1-sim2}  plots the true and estimated  partial functions for both intercept and
slope varying coefficients of the $\ln(\lambda)$ model, namely, $\beta_0(z_1, \overline{\bm{z}}_{-1}),\beta_1(z_1,\overline{\bm{z}}_{-1})$ where $\overline{\bm{z}}_{-1}$ is the vector of averages of all the moderator variables except $z_1$. We compared three methods: CMPBoost, CMPMOB  with split points estimated via exhaustive search, and CMPMOB  with split points estimated via 10\% change points.  The results illustrate that CMPBoost 
is able to reconstruct the underlying smooth function whereas CMPMOB provides only a piece-wise constant approximation. This is
a key advantage of using the CMPBoost model. Further, it is
computationally much simpler because of the base learner trees that are
fitted using ordinary linear regression trees (for a continuous variable) which are less computationally expensive. Similarly, Figure \ref{fig:boost-fn2-sim2}  plots the true and estimated  partial functions for both the intercept and
slope varying coefficients of the $\ln(\nu)$ model, namely, $\gamma_0(z_5, \overline{\bm{z}}_{-5}),\gamma_1(z_6,\overline{\bm{z}}_{-6})$. From the results in both figures, we see that all three methods under-estimated the $\ln \lambda$ model and over-estimated the $\ln \nu$ model. The reason is that the algorithms have terminated early 
 due to the non-smooth objective function surface. The complicated nature of the simulated data can be attributed to this finding. This was verified by conducting a less complicated simulation design that considers a linear function with no varying coefficient terms for the $\ln \nu$ model, 
  in which case the proposed methods estimated the true functions without any problems (see Appendix \ref{sec:boost-ex2}).

\begin{figure}[!htbp]
\centering
  \includegraphics[width=0.7\linewidth]{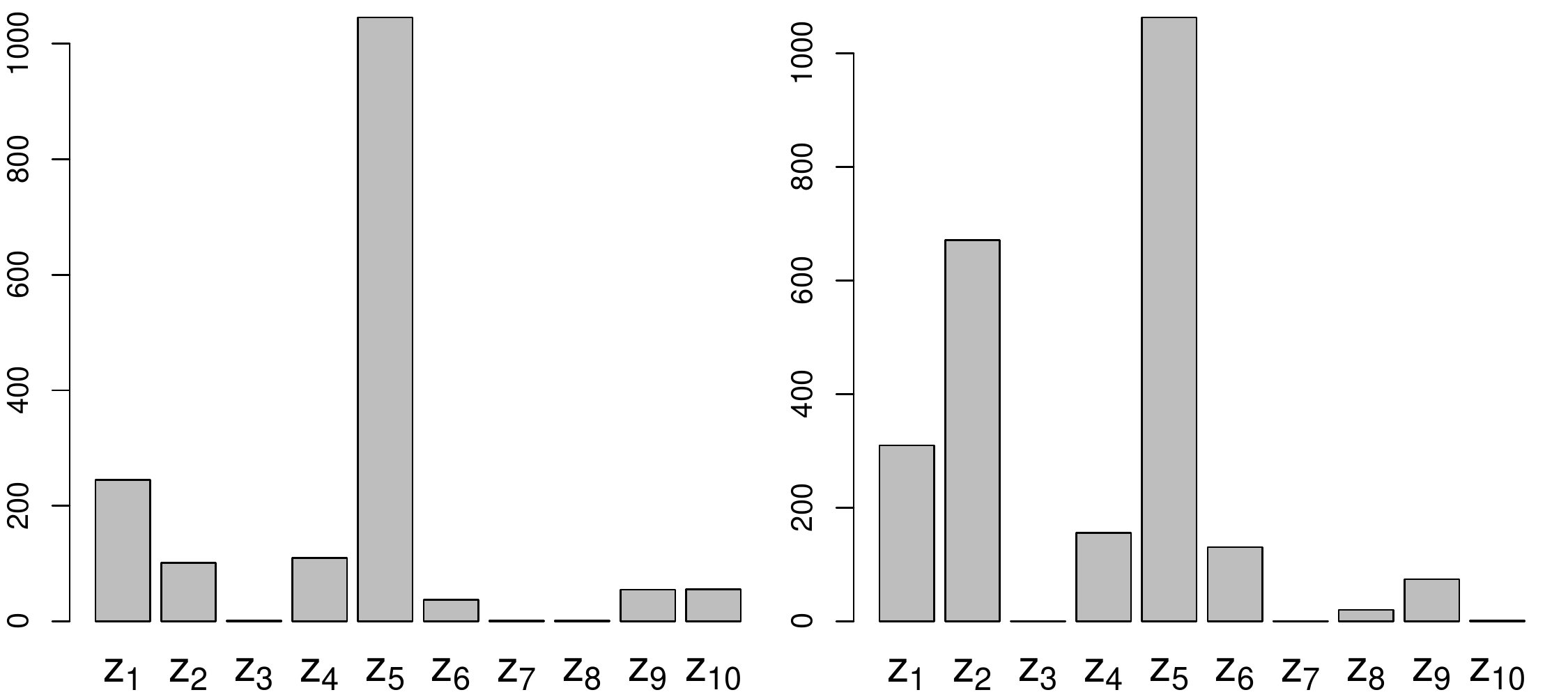}
\caption{Variable importance plots for the 10 moderator variables in boosting using  two different simulation datasets.}
\label{fig:boost-vi-sim2}
\end{figure}

The variable importance plots for two specific datasets among the 20 simulated datasets are described in Figure \ref{fig:boost-vi-sim2}. Although 
4 true moderator variables $(z_{i1},z_{i2},z_{i5},z_{i6})$ 
were used to simulate the data, most of the time the moderator variables $z_{i5}$ and $z_{i2}$ dominated the other moderator variables. This occurs 
even for the CMPMOB models. Hence, 
the variable importance plots are correctly identifying the dominating moderators.

On the whole, the results from this example  reiterate the fact that
CMPBoost 
is a more flexible and robust approach than CMPMOB. 
In the following section, we apply and compare CMPMOB and CMPBoost on 
a real data example.

%

\section{Modeling Data from a Bike Sharing System} \label{sec:mob-bikeshare}
We use CMPMOB and CMPBoost 
to analyze data from a bike sharing system. Bike sharing systems
are the new generation bike rental service providers. The
 data come from a bike sharing service provider in Washington DC \citep{,fanaee2014event}. For simplicity, we consider only data for January, 2012. It
 includes information on the number of hourly rentals 
 for registered and casual users, the weather,
whether or not the day is a holiday, etc. Table
\ref{tab:bikeshar-attr} provides the list of attributes and their descriptions. Descriptive statistics for the data are reported in Table \ref{tab:bike-descr} in Appendix \ref{apx: bik}. There are three missing values in the data and for that reason it included only 741 observations instead of 744 ($24 \times 31$) observations. In this study, we model only the casual
users; the same analysis can be repeated for the registered users.  
\begin{table}[h]
\small
	\caption{Full list of attributes and their description for the bike sharing data}
	\label{tab:bikeshar-attr}
		\centering
		\begin{tabular}{ll}
			\toprule
			Name & Description \\
			\midrule
			\emph{dteday} & date
\\
			\emph{season} & season (1:spring, 2:summer, 3:fall, 4:winter)
\\
			\emph{yr} & year (0: 2011, 1:2012)
\\
			\emph{mnth} & month ( 1 to 12)
\\
			\emph{hr} & hour (0 to 23)
\\
\emph{day} & day (1 to 30 or 31)
\\
			\emph{holiday} & whether or not the day is a holiday (extracted from\\ &\url{http://dchr.dc.gov/page/holiday-schedule}) \\
			\emph{weekday} & day of the week
\\
			\emph{workingday} & if day is neither weekend nor holiday is 1, otherwise is 0.
\\
			\emph{weathersit} & 1= Clear, Few clouds, Partly cloudy \\
				& 2= Mist + Cloudy, Mist + Broken clouds, Mist + Few clouds, Mist\\
				& 3= Light Snow, Light Rain + Thunderstorm + Scattered clouds \\
				& 4= Heavy Rain + Ice Pallets + Thunderstorm + Mist, Snow + Fog \\	
			\emph{temp} & Normalized temperature in Celsius. The values are divided to 41 (max)
\\
			\emph{atemp} & Normalized feeling temperature in Celsius. The values are divided to 50 (max)\\
			\emph{hum}& Normalized humidity. The values are divided to 100 (max)
\\
			\emph{windspeed} &Normalized wind speed. The values are divided to 67 (max)
\\
			\emph{casual} & count of casual users
\\
			\emph{registered} & count of registered users
\\
			\emph{cnt}& count of total rental bikes including both casual and registered \\
			\bottomrule
		\end{tabular}
	\end{table}

As a preliminary benchmark, we fit ordinary generalized linear models 
using both CMP and  Poisson regressions (see Table \ref{tab:glm-results} in Appendix \ref{apx: bik}). The CMP regression results in a large negative estimate for $\widehat{\gamma}_0$, indicating strong over-dispersion. 
The CMP and Poisson models diverge especially for variables 
\emph{hum} and \emph{windspeed}, which are highly significant in the Poisson but not in the CMP model.
 Not surprisingly, the CMP fit is much better 
 in terms of AIC and  log-likelihood. 
\subsection{Generalized Varying Coefficient Model}
Now, we consider the following varying coefficient model:
\begin{align}
  \begin{split}
    \ln \lambda &= \beta_0(windspeed,weathersit,holiday,weekday)\\ &+  atemp \times \beta_1(windspeed,weathersit,holiday,weekday) \\ & +hum \times \beta_2(windspeed,weathersit,holiday,weekday)\\&+hr \times \beta_3(windspeed,weathersit,holiday,weekday) \\ &+ day \times \beta_4(windspeed,weathersit,holiday,weekday),\\
    \ln \nu &= \gamma_0,
  \end{split}
  \label{eqn:tvc}
\end{align}
where  $\beta_1(\cdot), \beta_2(\cdot), \beta_3(\cdot)$ and $\beta_4(\cdot)$ are four-dimensional functions.
\begin{figure}[!h]
\centering
  \includegraphics[scale=0.5]{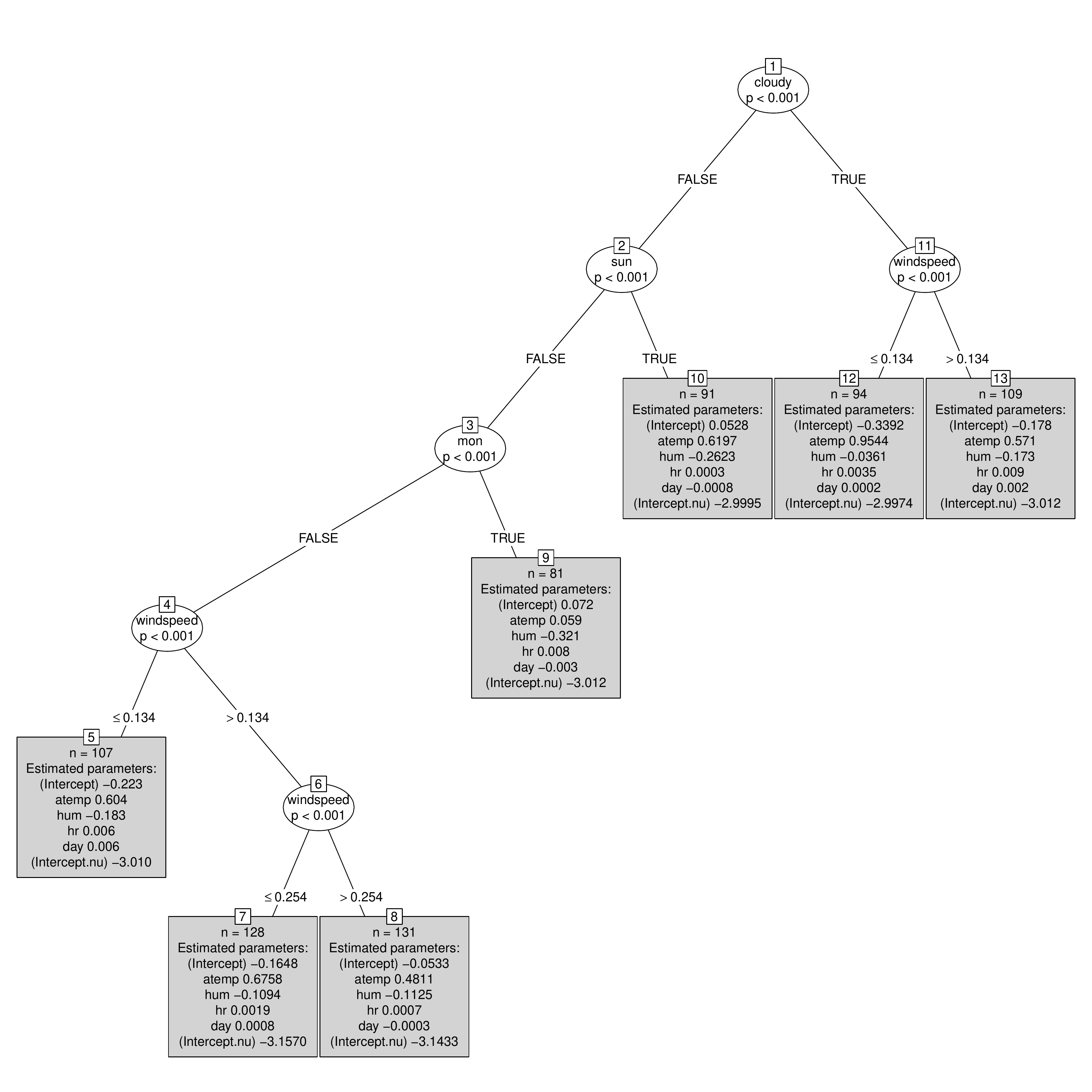}
    \caption{
    CMPMOB tree 
     with exhaustive search for (\ref{eqn:tvc}). 
    The local likelihood value $-2462.58$ is slightly lower than the global likelihood value without varying coefficients ($-2530.60$).}
  \label{fig:cmp-mob}
  \end{figure}

 Binary coding is employed for the categorical variables (\emph{weathersit}, 
 \emph{weekday}, 
 \emph{holiday}). 
 The tree for the estimated CMPMOB 
 with exhaustive search 
 is shown in Figure \ref{fig:cmp-mob}. The tree identified 
 \emph{cloudy}, \emph{windspeed}, \emph{sun} and
\emph{mon} as significant. The exact same tree is obtained for CMPMOB model with the 10\% change point method. 
The results indicate that ridership patterns are different for  cloudy vs. non cloudy days and on weekdays vs. weekends.  This is 
meaningful because 
weather and holidays play a significant role. Another important finding is that, although 
\emph{windspeed} is not significant in the CMP regression model as a main effect, it is 
significant as a moderator in the CMPMOB tree.

The estimated Poisson MOB 
(Figure \ref{fig:pos-mob} in Appendix \ref{apx: bik}) 
identified  \emph{sat} and
\emph{sun} as significant moderators. Surprisingly, 
\emph{windspeed} which is selected as a main effect in the Poisson regression is not included in the tree. 
This might be due to the inability of the Poisson distribution to capture over-dispersion. 

\begin{figure}[h]
 \centering
  \includegraphics[height=0.35\textheight]{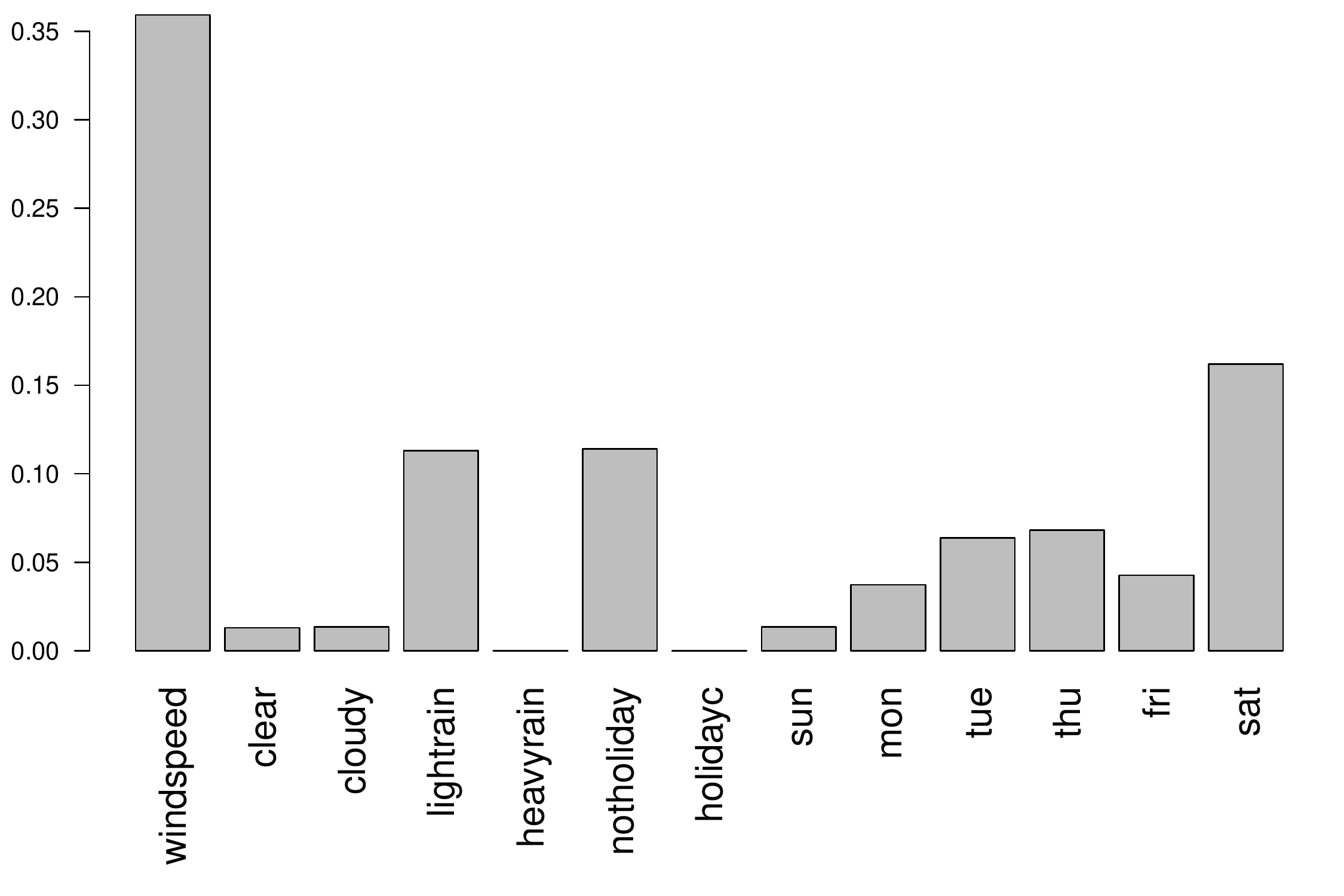}
  \caption{Variable importance plot for CMPBoost} 
  \label{fig:vi-bike}
\end{figure}

To assess the variable importance and to provide smoother approximations for Model (\ref{eqn:tvc}), we fitted the CMPBoost model. The tuning parameter for minimum node size is chosen as 20, and $\xi=0.8$ is chosen to reduce the number of boosting iterations $B$, which is determined automatically. For the number of terminal nodes $M$, we used a test data (January, 2011) to find the optimal value $M=25$. With these tuning parameters, the CMPBoost took $B=1252$ iterations. The variable importance plot is shown in Figure \ref{fig:vi-bike}. Similar to the CMPMOB tree, the CMPBoost 
identified 
\emph{windspeed} as the most significant. Apart from that, 
\emph{sat}, \emph{lightrain} and \emph{notholiday} have high importance scores. Surprisingly, \emph{cloudy}, which is selected as a significant variable in the CMPMOB Tree in Figure \ref{fig:cmp-mob}, has a low variable importance score. This could be attributed to the instability problem of a single tree. 

\begin{figure}[h]
  \includegraphics[width=1\linewidth]{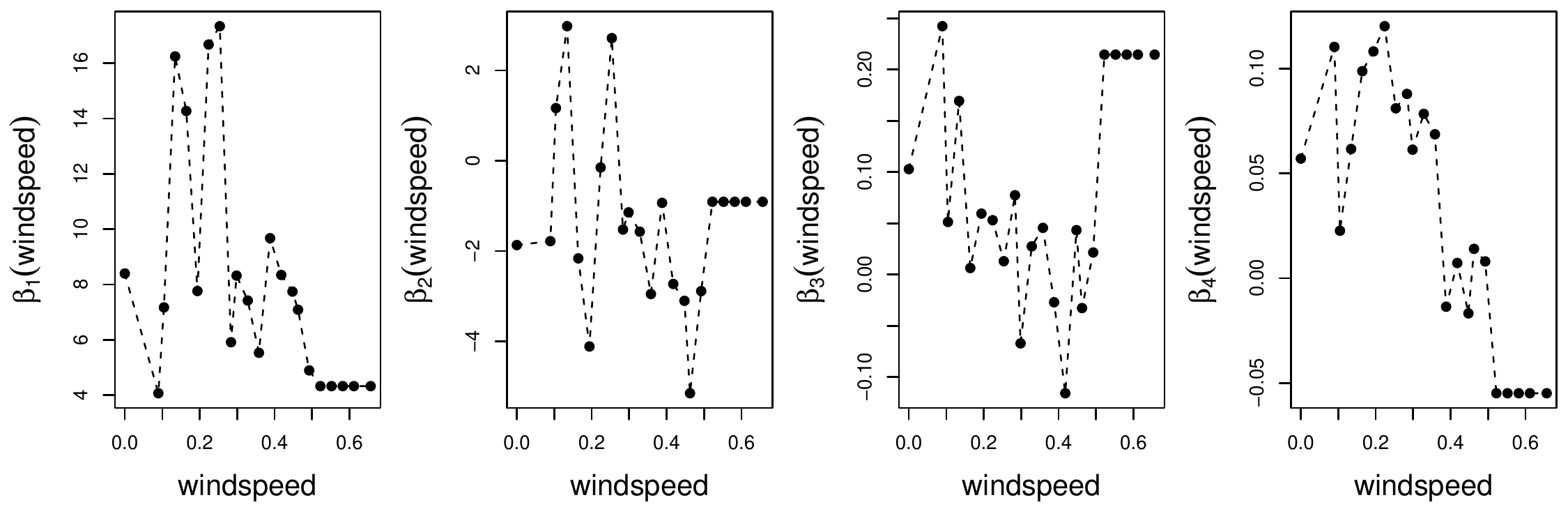}
  \caption{partial dependence plot for variable \emph{windspeed} in  CMPBoost}
  \label{fig:part1-bike}
\end{figure}

To gain further insight into the results from the CMPBoost model, partial dependence plots for 
\emph{windspeed} are constructed (Figure \ref{fig:part1-bike}). Similar plots can be constructed for other significant moderator variables such as \emph{sat}, \emph{notholiday} or \emph{lightrain}.  Figure \ref{fig:part1-bike} shows that for moderate wind speed, 
\emph{atemp} seems to have a 
positive effect on ridership. 
Similarly, 
\emph{hum} seems to have an overall negative effect on ridership 
except for a few moderate wind speed values. In contrast, 
\emph{hr} 
has a positive relationship with ridership 
except for a few moderate values of wind speed.  It is surprising to see that the partial dependence plots in Figure \ref{fig:part1-bike} are not smooth as expected in the CMPBoost model. This may be due to the presence of categorical variables in the functions in (\ref{eqn:tvc}) that make the underlying true function itself not smooth. 

\subsection{A More Flexible Generalized Semi-Varying Coefficient Model}
\begin{figure}[h]
	\centering
	\vspace{-0.2in}
	\includegraphics[scale=0.6]{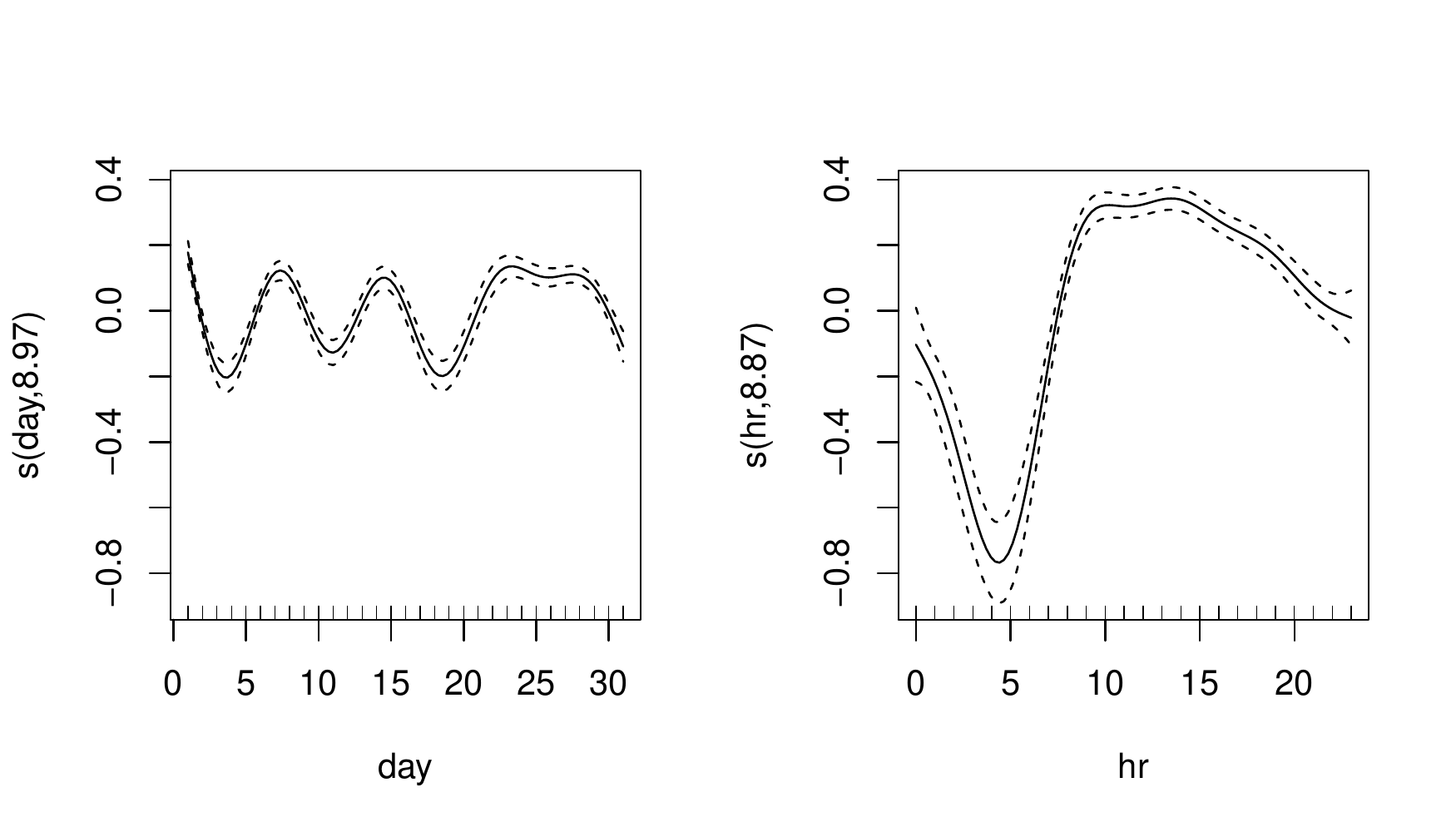}
    \caption{The fitted global smooth functions $f_1(\cdot)$ and $f_2(\cdot)$ for the CMPMOB 
    model in (\ref{eqn:tvc-gen}). $s(\cdot)$ are the estimated smooth functions. CMPBoost yields the same result.}
    \label{fig:cmp-glob-smooth}
\end{figure}

To provide even more flexibility, 
we consider the following semi-varying coefficient model:
\begin{align}
  \begin{split}
    \ln \lambda  &= \beta_0(weathersit,holiday,weekday) \\ &+ atemp \times \beta_1(weathersit,holiday,weekday) + f_1(day) + f_2(hr) \\ 
    \ln \nu &= \gamma_0(weathersit,holiday,weekday)\\ &+windspeed \times \gamma_1(weathersit,holiday,weekday)+ \gamma_2 hum,
  \end{split}
  \label{eqn:tvc-gen}
\end{align}
where  $\gamma_2$ and  functions $f_1(\cdot)$, $f_2(\cdot)$ are global, and are not re-estimated at each
node of the tree or in the boosting algorithm. The tree for the CMPMOB 
(Figure \ref{fig:cmp-mob-gen} in Appendix \ref{apx: bik}) 
selected the variables \emph{sat}, \emph{fri}, \emph{mon}, \emph{clear} and \emph{cloudy} as significant moderators. The variable \emph{atemp} has a positive effect on ridership 
except on Mondays, when the effect seems to be negative. Similarly, \emph{windspeed} seems to create 
more variation in ridership 
on Fridays, Saturdays and Mondays if the weather is clear, and also on the remaining days if the weather is cloudy. The global smooth functions $f_1(\cdot), f_2(\cdot)$ are presented in
Figure \ref{fig:cmp-glob-smooth}, showing that
rentals 
follow a cyclical pattern across  days, which
could be  due to a weekend/weekday effect. Further, rentals are highest in the afternoon  compared to evenings or early mornings.  

\begin{figure}[h]
	\centering
	\includegraphics[width=1\textwidth]{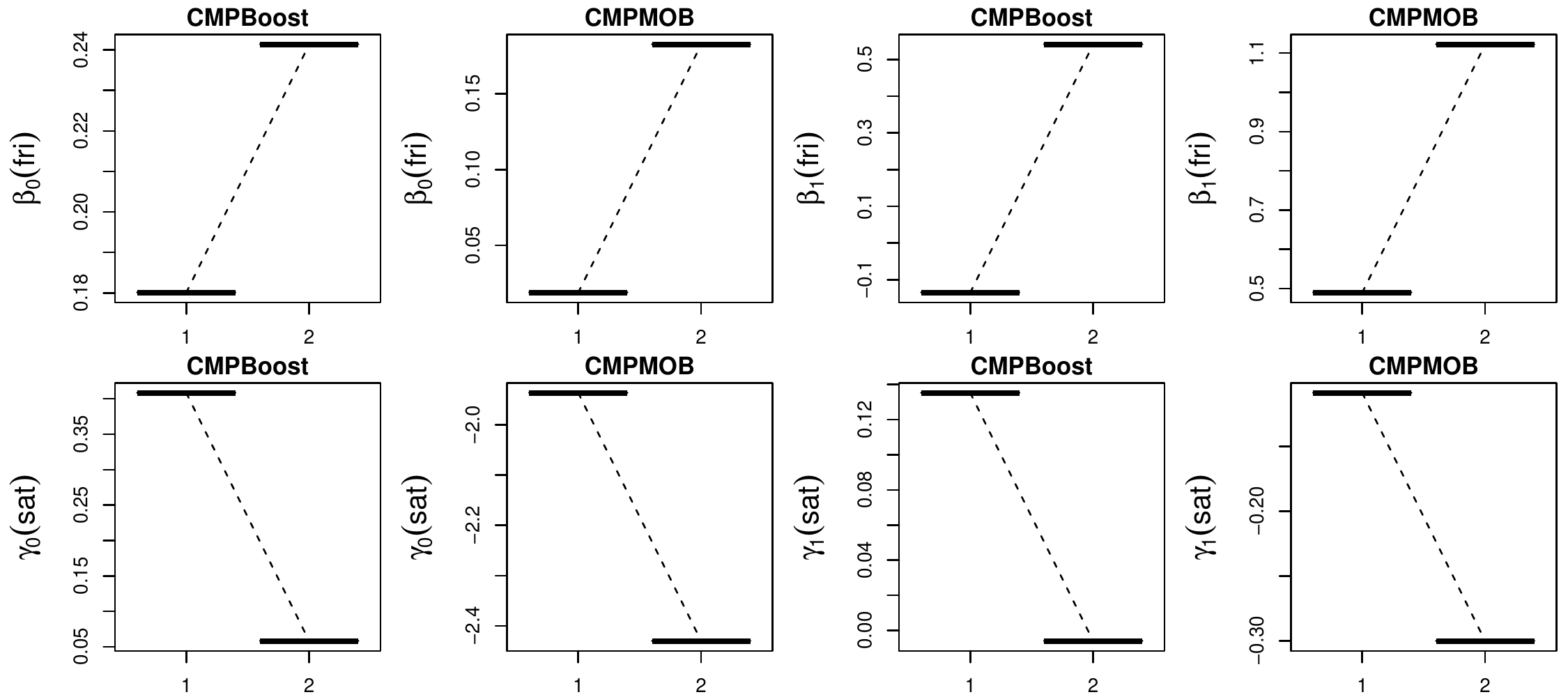}
    \caption{partial dependence plots for the  $\beta_0(\cdot)$, $\beta_1(\cdot)$, $\gamma_0(\cdot)$ and $\gamma_1(\cdot)$ functions   of (\ref{eqn:tvc-gen}), estimated in CMPBoost and CMPMOB models. Here, 1=FALSE, 2=TRUE. The top panel is for \emph{fri} and the bottom panel is for \emph{sat}.}
    \label{fig:part-gen}
\end{figure}

The CMPBoost model is fitted using the same tuning parameters as  
in the previous model. The overall likelihood value ($-1917.26$) is slightly better than the CMPMOB tree. Comparing the CMPMOB tree (Figure \ref{fig:cmp-mob-gen} in Appendix \ref{apx: bik}) and the variable importance measures of the CMPBoost 
(Figure \ref{fig:boost-vi-gen} in Appendix \ref{apx: bik}), 
in both cases \emph{fri} and \emph{sat} are the most influential variables. More importantly, the $\ln \lambda$ model selected 
\emph{fri} as the most influential moderator whereas the $\ln \nu$ model selected the variable \emph{sat}. Based on this finding, we constructed the partial dependence plots for the coefficient functions $\beta_0(\cdot)$, $\beta_1(\cdot)$ with respect to 
\emph{fri}, and for the functions $\gamma_0(\cdot)$, $\gamma_1(\cdot)$ with respect to 
\emph{sat}. The results are shown in Figure \ref{fig:part-gen}.

To conclude, the results show that both the CMPBoost and CMPMOB 
are very flexible for modeling the bike sharing data and they provide 
valuable insights.

\section{Summary and Conclusions} \label{sec:mob-summary}

We proposed a novel tree-based semi-varying coefficient model for the CMP distribution. Our CMPMOB model formulation is 
more flexible  than 
existing tree-based models for count data, as it allows including node-invariant (global) effects in the model. A known drawback of tree-based methods is that they only provide a piece-wise constant approximation to the underlying smooth functions. To overcome this limitation, we also developed a boosting approach, CMPBoost, to provide  smoother estimates of the underlying varying coefficient functions. We provide R code for all of these procedures at \url{https://github.com/SuneelChatla/CMPTree}. 

We used the MOB algorithm to estimate the proposed tree-based semi-varying coefficient model because it provides  coefficient-constancy tests that are
easy to implement. The methodology can be extended to any tree-based estimation approach, such as PartReg. The existing MOB algorithm uses exhaustive search to identify  potential split points and is computationally intensive for the CMP distribution. 
We provided some heuristics to reduce the
computational burden 
and at the same time, we  proposed a new split point estimation approach by borrowing tools from the change-point estimation literature.  

On the whole, the proposed semi-varying coefficient models (CMPBoost and CMPMOB) are shown to be useful 
for handling count data. In addition to their ability to capture both over- and under-dispersion, their flexibility to model complex nonlinear relationships makes them powerful 
tools for analyzing count data in a wide range of 
applications.

\singlespacing
\bibliographystyle{agsm}
\bibliography{references}
\newpage

\renewcommand{\thesection}{A\arabic{section}}
\setcounter{subsection}{0}
\setcounter{section}{0}

\renewcommand{\thetable}{A\arabic{table}}
\renewcommand{\thefigure}{A\arabic{figure}}
\setcounter{figure}{0}
\setcounter{table}{0}

\section{Appendix A1: CMPMOB Algorithm}
\label{apx:algo}
   The algorithm for CMPMOB semi-varying coefficient model is given below. First, it fits a global model with all the terms in the model including fixed effects terms using all the data. Once the fixed/global effects are estimated, the MOB procedure starts by constructing a tree with the splitting variables. The estimated fixed effects values will be treated as offset terms in all the subsequent models. After the tree is fitted, the fixed/global effects are re-estimated using the tree model as an offset term.
   \begin{algorithm}
   \scriptsize
 	\DontPrintSemicolon
 	\caption{Estimation procedure for CMPMOB semi-varying coefficient model.}\label{alg:mob-cmp} \vspace{0.2in}
 	\KwIn{
 		Initialize $\mathcal{B}_1 \leftarrow \mathcal{Z}_1 \times \cdots \times \mathcal{Z}_L $ and $M \leftarrow 1$.  \;
 		Obtain consistent estimates for $\bm{\phi}_1, \bm{\phi}_2$ by fitting a model on the entire data. }
 	
 	Fit the CMP model with  predictor functions
 	\begin{eqnarray*}
 		\eta_{1i} = \sum_{m=1}^{M} 1(\bm{z}_i \in \mathcal{B}_m)  \left(\bm{x}_{1i}^{T}\bm{\beta}_m+ \bm{x}_{2i}^{T}\widehat{\bm{\phi}}_1 \right), \quad
 		\eta_{2i} =\sum_{m=1}^{M} 1(\bm{z}_i \in \mathcal{B}_m)  \left(\bm{w}_{1i}^{T}\bm{\gamma}_m + \bm{w}_{2i}^{T} \widehat{\bm{\phi}}_2 \right). 
 	\end{eqnarray*} \;
 	Apply tests for constancy of the coefficients $\bm{\beta}$ and $\bm{\gamma}$ separately for each $Z_l$, $l=1,\ldots, L$. This yields $L \times 2$, $p-$values, $p_{11},\ldots,p_{L2}$ \;
 	\If{$p_{min} := \text{min}(p_{11},\ldots,p_{L2})$}{
 		Select the variable $Z_l$ and node $\mathcal{B}_s$ where $p_{l}=p_{min}$ \;
 		\If{change point=TRUE}{
 			Sort the score functions $\bm{x}_{1i}(y_i-\widehat{E[y_i]})$ and $\bm{w}_{1i} (-\ln(y_i!)+ \widehat{E[\ln(y_i!)]})\nu_i$ by $Z_l$\;
 			Estimate the  test statistic $(\widehat{\ell}_{\Delta_k})$ values assuming that change in mean or variance occurs at value $k$\;
 			For exact search, identify the $k$ values for all the best statistics and for percentage search, identify the  $k$ values for best 5\% or 10\%.
 		}
 		\Else{
 			\ForEach{unique candidate split $\Delta_k$ in $\{z_{li}: \bm{z}_{i} \in \mathcal{B}_s \}$ that divide $\mathcal{B}_s$ into two nodes $\mathcal{B}_{sk1}$ and $\mathcal{B}_{sk2}$}{Compute $\widehat{\ell}_{\Delta_k}= \text{max} \quad \ell_{\Delta_k}(\bm{\beta}, \bm{\gamma})$ of the CMP model 
 				\begin{eqnarray*}
 					\eta_{1i}=\sum_{m=1}^{2} 1(\bm{z}_i \in \mathcal{B}_{skm}) \left(\bm{x}_{1i}^{T}\bm{\beta}_{km}+ \bm{x}_{2i}^{T} \widehat{ \bm{\phi}}_1\right),
 					\eta_{2i}=\sum_{m=1}^{2} 1(\bm{z}_i \in \mathcal{B}_{skm})  \left(\bm{w}_{1i}^{T}\bm{\gamma}_{km}+ \bm{w}_{2i}^{T} \widehat{\bm{\phi}}_2\right). 
 			\end{eqnarray*}}
  }
 		Split $\mathcal{B}_s$ into $\mathcal{B}_{s1}, \mathcal{B}_{s2}$ by $\widehat{\Delta}_k = \text{arg max}_{\Delta_k}\widehat{\ell}_{\Delta_k} $	 and set $ M \leftarrow M+1$\;
 		\For{kid in 1:2}{
 			go to step 1 with $\{\bm{z}_{i} \in \mathcal{B}_{skid} \}$
 		}
 	}
 	Re-estimate $\bm{\phi}_1, \bm{\phi}_2$ using the following model
 	\begin{eqnarray*}
 		\eta_{1i} = \sum_{m=1}^{M} 1(\bm{z}_i \in \mathcal{B}_m)  \bm{x}_{1i}^{T}\widehat{\bm{\beta}}_m+ \bm{x}_{2i}^{T}\bm{\phi}_1, \quad
 		\eta_{2i} =\sum_{m=1}^{M} 1(\bm{z}_i \in \mathcal{B}_m)  \bm{w}_{1i}^{T}\widehat{\bm{\gamma}}_m + \bm{w}_{2i}^{T} \bm{\phi}_2. 
 	\end{eqnarray*}
 	
 	
 \end{algorithm}

   
   
   %

 \section{Appendix A2: CMPMOB Simulation with Different Split Variables for the $\ln(\nu)$ model}

\label{sec:mob-ex2}
 The simulation design for this example is same as for the example in Section \ref{sec:mob-sim} except that both $\ln(\lambda)$ and $\ln(\nu)$ use different split variables and split points. More specifically,  we now consider $\eta_1=2+1(z_1>0.65)2x_1+1(z_1\le 0.65)x_2+2f_1^2(x_3)$ with $\lambda = \exp(\eta_1)$ and $\eta_2=0.25+1(z_3>0.5)0.5w_1+0.5f_2^2(w_2)$ with $\nu = \exp(\eta_2)$. The remaining details are same as in Section \ref{sec:mob-sim}.
 
\begin{figure}[!htbp]
     \centering
     \begin{minipage}{0.45\textwidth}
     \includegraphics[width=1.2\textwidth]{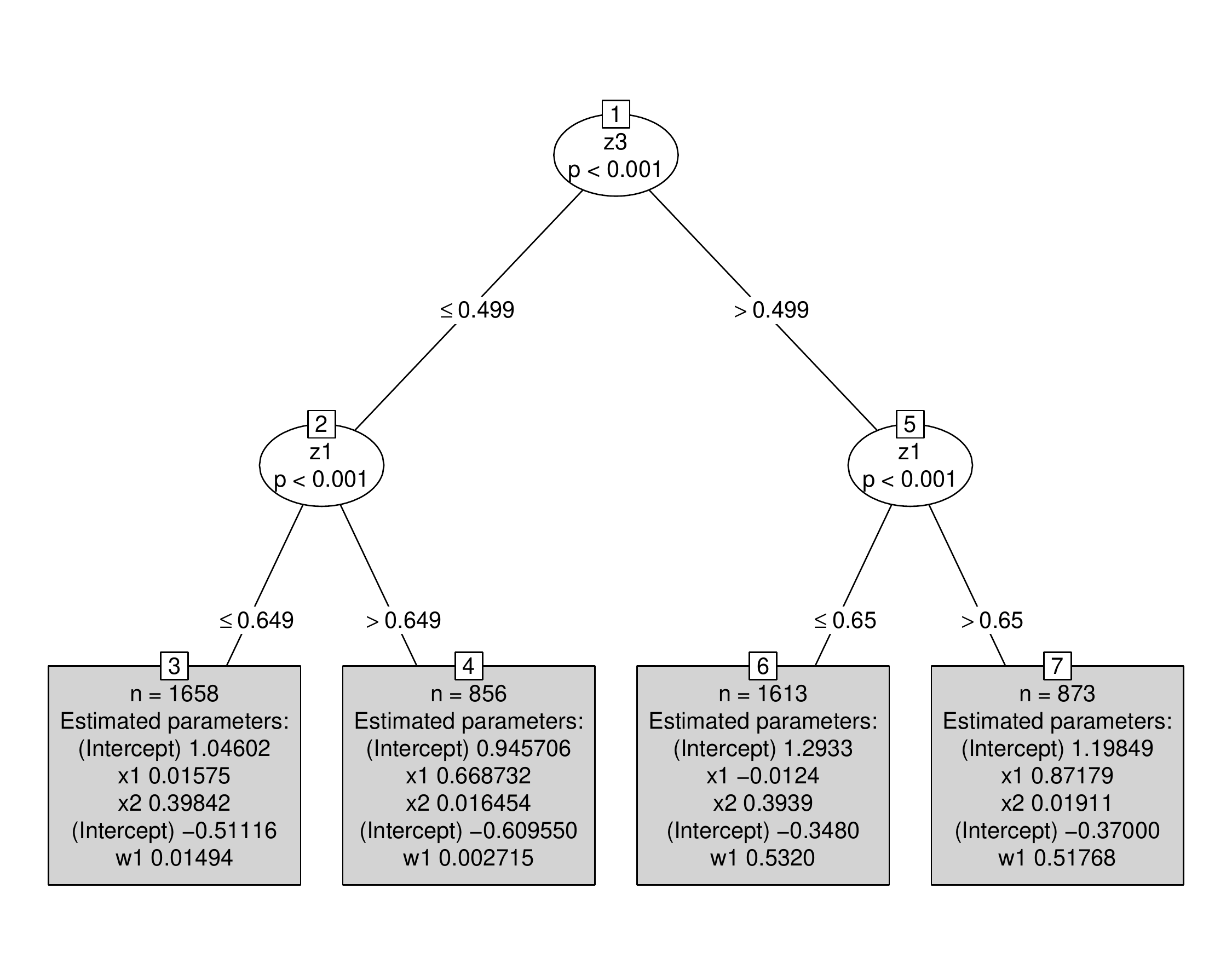}
     \end{minipage}
     \begin{minipage}{0.45\textwidth}
     \includegraphics[width=1.2\textwidth]{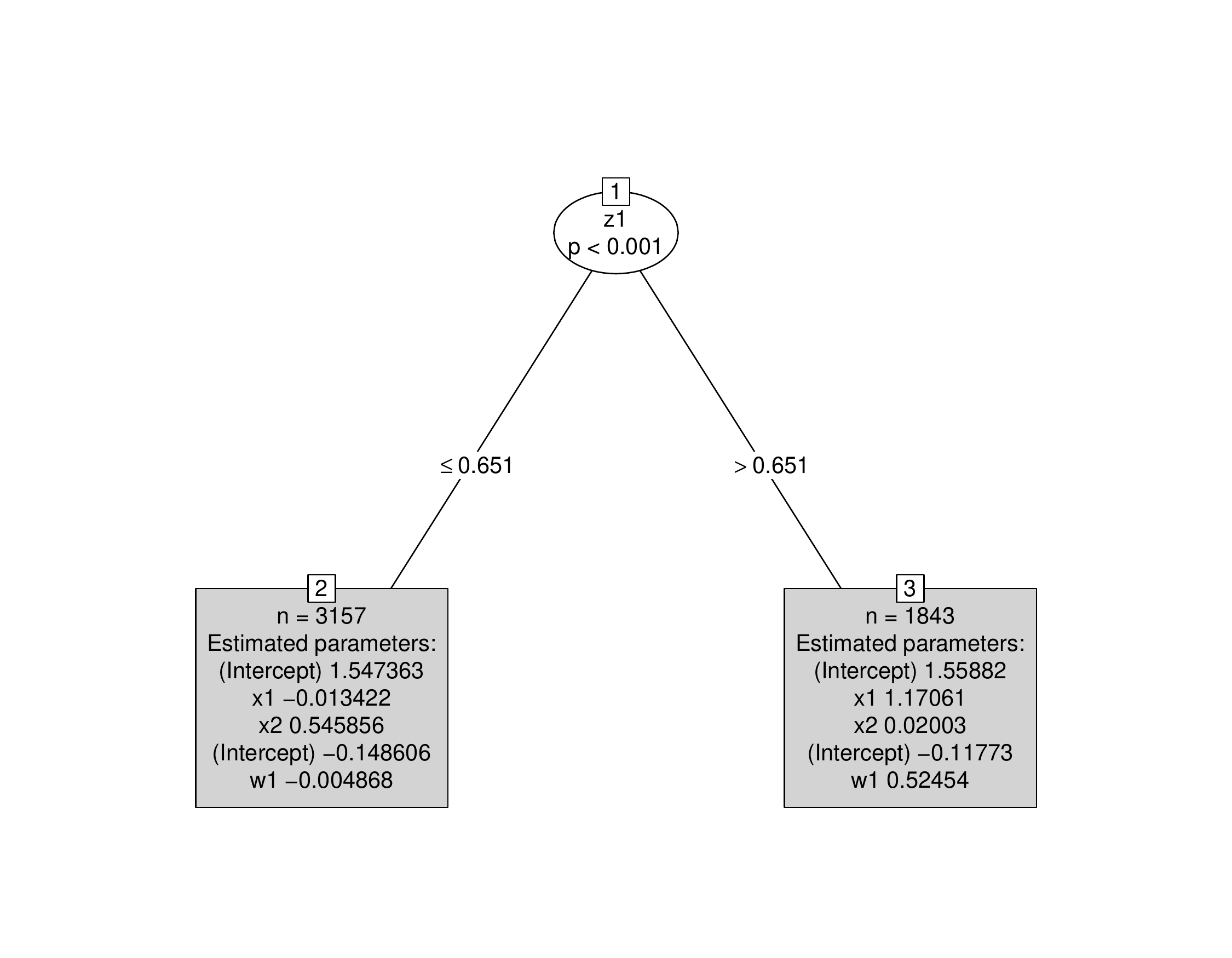}
     \end{minipage}
     \caption{Estimated trees for two data sets (repetition 20 and 1) with $n=5000$. Tree structures are similar for the remaining data sets. Fixed effects plot is omitted for brevity.}
     \label{fig:sim1-ex2-tree}
 \end{figure}

\begin{figure}[!htbp]
     \centering
     \includegraphics[scale=0.5]{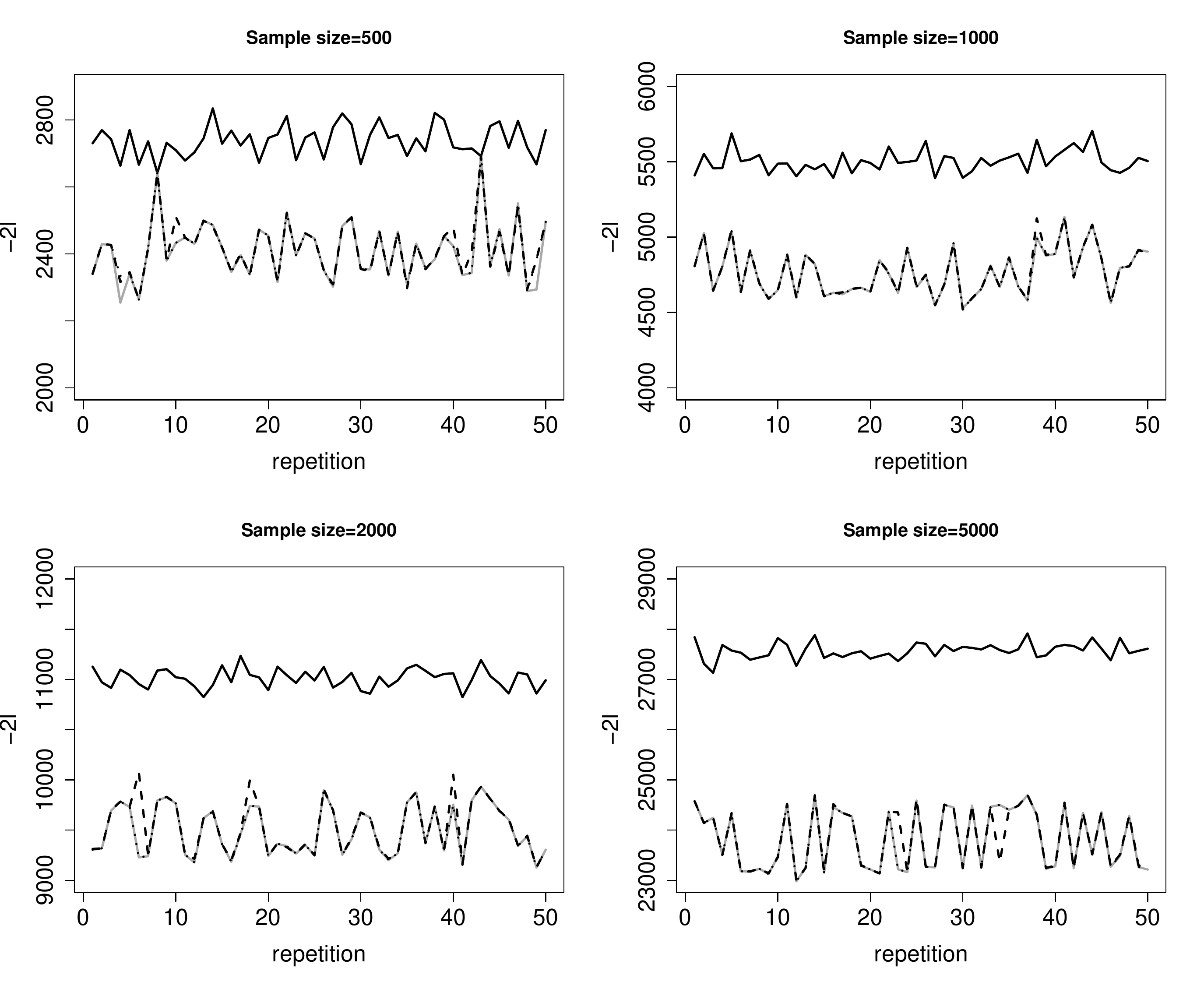}
     \caption{$-2l$ values for the three models, for each sample size, and across 50 replications. (--- global , {\color{gray}---} exhaustive search, - - exact change points, .... 10\% change points)}
     \label{fig:sim1-ex2-dev}
 \end{figure}

 The estimated tree for one of the data sets is shown in Figure \ref{fig:sim1-ex2-tree}. This time, the first  tree chose  4 terminal nodes to accommodate the two true underlying splits. The estimated splits are $(\{z_3\le 0.5, z_1 \le 0.65\}, \{z_3\le 0.5, z_1 > 0.65 \} ; \{z_3> 0.5, z_1 \le 0.65\}, \{z_3> 0.5, z_1 > 0.65\})$. On the other hand, the second tree has only 3 terminal nodes in which the variable $z_1$ is probably not significant under $z_3\le 0.5$. Obviously, the fit measure for the first tree (rep 1) is much better than the second tree (rep 20) which is evident from the right bottom plot in Figure \ref{fig:sim1-ex2-dev}.  Note that both $\ln(\lambda)$ and $\ln(\nu)$ must be estimated simultaneously using the same data and hence they cannot have different splits at the same time. This is the reason why we considered the same splitting variables in both (\ref{eqn:vc1}) and (\ref{eqn:vc2}).

 The findings from this example are consistent with the findings in Section \ref{sec:gb-sim}. As shown in Figure \ref{fig:sim1-ex2-dev} and Table \ref{tab:mob-sim-wglo-ex2}, the results from 10\% change points are identical to the results from exhaustive search except for the data sets with  $n=500$. Even in that case, the results are not practically different as their overall fit measures are very close. The results from the model estimated using the exact set of change points are also very close to the results from the model estimated using exhaustive search, both in terms of overall fit 
 and the estimated split points.

  \begin{sidewaystable}[!htbp]
  \footnotesize
  \centering
  	\begin{tabular}{ccccc}
  		\hline
  		
  		& \multicolumn{4}{c}{Exhaustive Search} \\
  		\cline{2-5}  
  		& $n=500$ & $n=1000$ & $n=2000$ &$n=5000$ \\
  		\hline 
  		Split-1& $z_3: 0.50(0.007), z_1: 0.647(0.0007)$  & $z_3: 0.499(0.003), z_1: 0.65(n.a)$ & $z_3: 0.499(0.002), z_1: 0.649(n.a)$ & $z_3: 0.500(0.0007), z_1: 0.651(n.a)$  \\
  		Split-2& $z_1: 0.641(0.016), z_3:0.51(0.013)$  & $ z_1: 0.644(0.019), z_3: 0.49(n.a)$ & $z_1: 0.648(0.004), z_3: 0.499(n.a)$ & $z_1: 0.649(0.002), z_3: 0.50(n.a)$  \\
  		Split-3& $z_1:0.643(0.027)$  & $ z_1: 0.648(0.011), z_3: 0.415(0.127)$ & $z_1: 0.626(0.115), z_3: 0.499(n.a)$ & $z_1: 0.649(0.002), z_3: 0.501(n.a)$  \\
  		Split-4& $--$  & $ z_1: 0.66(n.a)$ & $z_1: 0.648(n.a)$ & $--$  \\
  		
  		Global $-2l$ & $2738.69(47.52)$   & $5504.18 (73.89)$  & $11010.67 (95.49)$ & $27573.11 (159.45)$  \\ 
  		Local $-2l$ & $2410.39 (89.28)$   & $4768.56 (154.19)$  & $ 9501.31 (246.51)$ & $23818.77 (607.49)$  \\
  		\hline 
  		
  	\end{tabular} 
  	\begin{tabular}{ccccc}
  		\hline
  		&  \multicolumn{4}{c}{Search with Exact Change Points} \\
  		\cline{2-5}  
  		& $n=500$ & $n=1000$ & $n=2000$ &$n=5000$  \\
  		\hline 
  		Split-1&  $z_3: 0.498(0.005), z_1: 0.65(0.0026)$  & $ z_3: 0.498(0.003), z_1: 0.650(n.a)$ & $z_3: 0.499(0.002), z_1: 0.67(n.a)$ & $z_3: 0.499(0.001), z_1: 0.651(n.a)$  \\ 
  		Split-2&  $z_1: 0.65(0.015), z_3: 0.50(0.002)$  & $ z_1: 0.65(0.01), z_3: 0.498(n.a)$ & $z_1: 0.652(0.007), z_3: 0.499(n.a)$ & $z_1: 0.652(0.004), z_3: 0.499(n.a)$  \\ 
  		Split-3&  $z_1: 0.645(0.037), z_3: 0.715(n.a)$  & $ z_1: 0.654(0.02), z_3: 0.491(n.a)$ & $z_1: 0.652(0.008), z_3: 0.498(n.a)$ & $z_1: 0.651(0.004), z_3: 0.50(n.a)$  \\ 
  		
  	Global $-2l$ & $2738.69(47.52)$   & $5504.18 (73.89)$  & $11010.67 (95.49)$ & $27573.11 (159.45)$  \\
  		Local $-2l$ & $2416.85 (86.22)$   & $4772.21 (159.30)$  & $9532.88 (269.69)$  & $ 23825.41 (602.68)$  \\
  		\hline 
  		
  	\end{tabular}
  	\begin{tabular}{ccccc}
  		\hline
  		&  \multicolumn{4}{c}{Search with 10\% Change Points} \\
  		\cline{2-5}  
  		& $n=500$ & $n=1000$ & $n=2000$ &$n=5000$  \\
  	\hline
  		Split-1& $z_3: 0.499(0.006), z_1: 0.647(0.001)$  & $z_3: 0.499(0.003), z_1: 0.65(n.a)$ & $z_3: 0.499(0.001), z_1: 0.649(n.a)$ & $z_3: 0.500(0.0006), z_1: 0.650(n.a)$  \\
  		Split-2& $z_1: 0.64(0.21), z_3:0.51(0.013)$  & $ z_1: 0.644(0.02), z_3: 0.49(n.a)$ & $z_1: 0.648(0.004), z_3: 0.499(n.a)$ & $z_1: 0.649(0.002), z_3: 0.499(n.a)$  \\
  		Split-3& $z_1:0.645(0.02)$  & $ z_1: 0.647(0.011), z_3: 0.415(0.127)$ & $z_1: 0.625(0.11), z_3: 0.499(n.a)$ & $z_1: 0.649(0.002), z_3: 0.50(n.a)$  \\
  		Split-4& $--$  & $ z_1: 0.66(n.a)$ & $z_1: 0.648(n.a)$ & $--$  \\
  		
  		Global $-2l$ & $2738.69(47.52)$   & $5504.18 (73.89)$  & $11010.67 (95.49)$ & $27573.11 (159.45)$  \\ 
  		Local $-2l$ & $2413.74 (85.57)$   & $4768.56 (154.19)$  & $ 9501.31 (246.51)$ & $23818.77 (607.49)$  \\
  		\hline 
  		
  	\end{tabular}
  	\caption{\label{tab:mob-sim-wglo-ex2} Model  $\ln(\lambda) \sim
      \beta_0(z_1,z_2,z_3,z_4)+ \beta_1(z_1,z_2,z_3,z_4)x_1+\beta_2(z_1,z_2,z_3,z_4)x_2+
      s(x_3); \ln(\nu) \sim  \gamma_0(z_1,z_2,z_3,z_4)+ \gamma_1(z_1,z_2,z_3,z_4)w_1+
      s(w_2) $ for the example 2 data sets with the smooth $s(x_3), s(w_2)$ as fixed effects. Values in  parenthesis are standard deviations for $50$ simulations and ($n.a$ means there is only one observation and hence standard deviations are not computed).} 
  \end{sidewaystable}

\clearpage 
 
\section{Appendix A3: CMPBoost Simulation with Linear Function (no varying coefficient terms) for $\ln(\nu)$}
\label{sec:boost-ex2}

The simulation design is exactly as described in Section \ref{sec:gb-sim} except that we do not use varying coefficients for the $\ln(\nu)$ model, but rather a linear function of the form $\nu=\text{exp}(0.25+0.25w_i))$.

\begin{figure}[h]
    \centering
    \vspace{-0.2in}
    \includegraphics[width=0.5\textwidth]{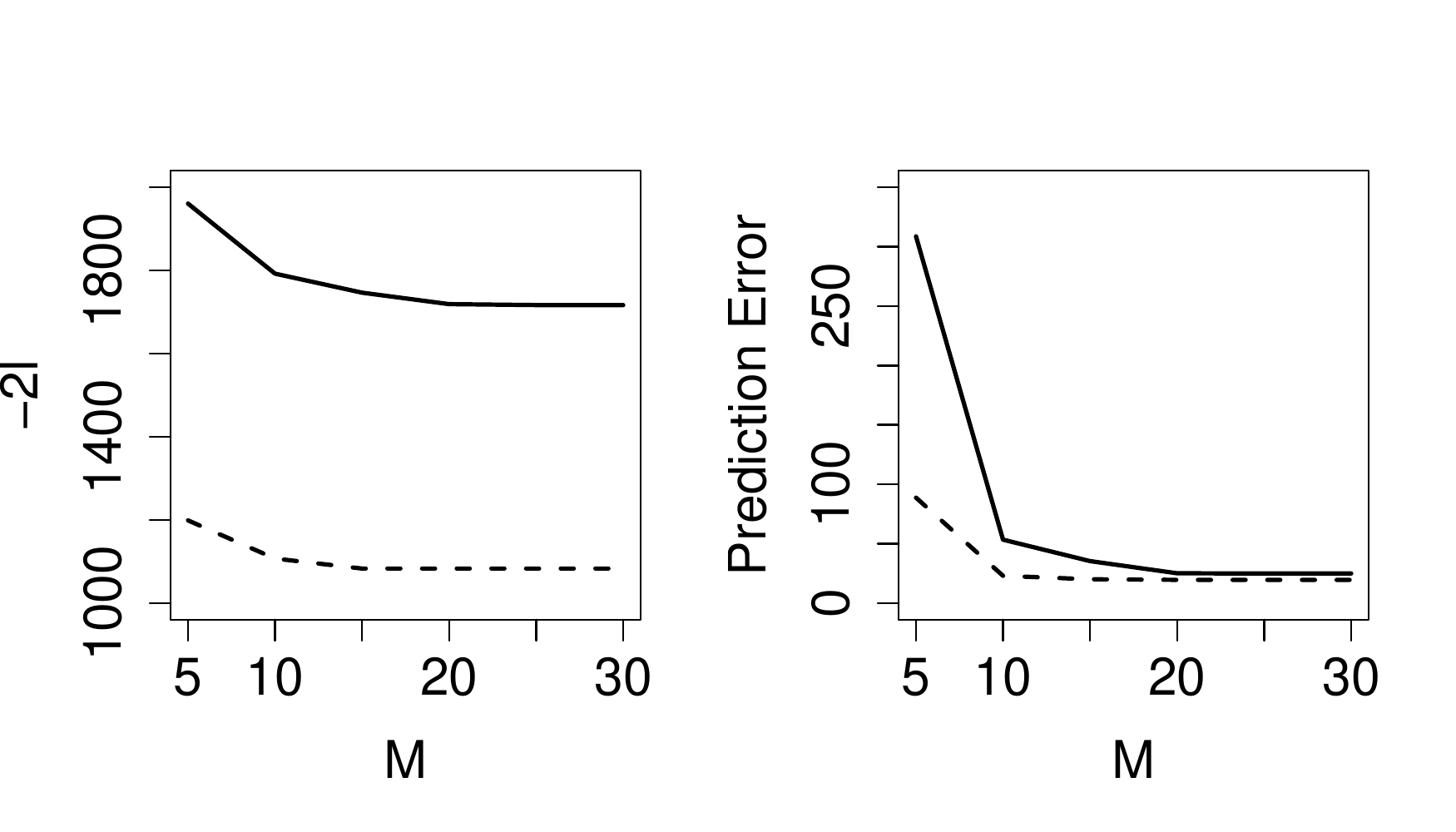}
    \caption{\textbf{Left:} -2 log-likelihood values. \textbf{Right:} prediction error for the models with an increasing number of terminal nodes ($M$). Solid line represents the values for the training data and dashed line for the test data. In both plots, the minimum for test data is obtained at $M=15$ and 500 iterations are used with $\xi=0.1$ for each model.}
    \label{fig:gb-train-m}
\end{figure}

\begin{figure}
\centering
  \includegraphics[width=0.8\textwidth]{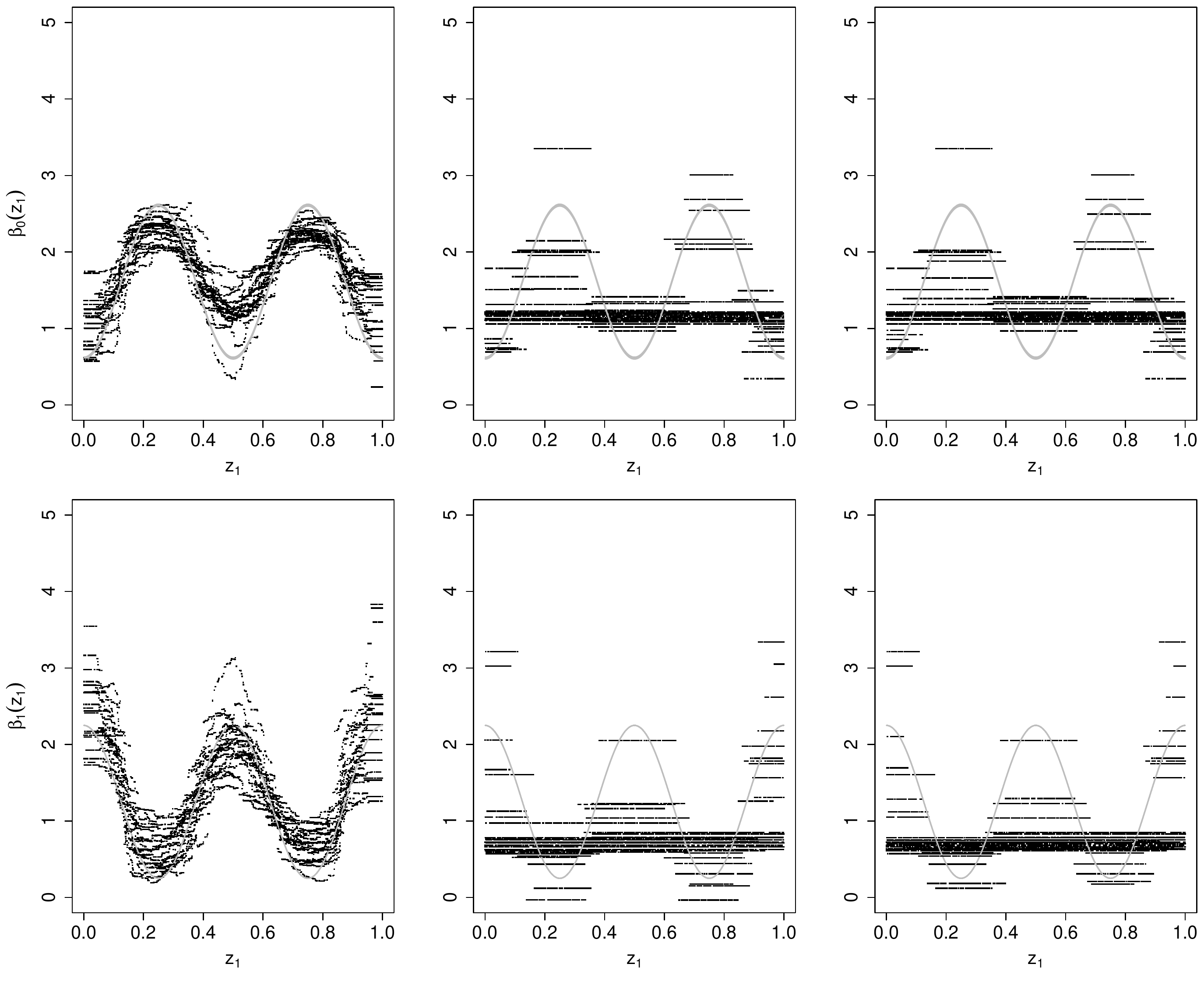}
\caption{\textbf{Top:} Reconstructed varying coefficient surfaces for $\beta_0(\bm{z})$. and \textbf{Bottom:} $\beta_1(\bm{z})$ for boosting, MOB tree fitted with exhaustive search, and MOB tree fitted with change point (10\% points), using 20 simulations.}
\label{fig:boost-fn1}
\end{figure}

Figure \ref{fig:gb-train-m} shows the results for the models fitted with an increasing number of terminal nodes for base learners ($M$) on both training and test data. 
Both $-2l$ and prediction error have similar behavior and obtain their minimum at $M=15$ on the test data.
Figure
\ref{fig:boost-fn1}  plots the true and estimated  partial functions for both intercept and
slope varying coefficients $(\beta_0(z_1, \overline{\bm{z}}_{-1}),\beta_1(z_1,\overline{\bm{z}}_{-1}))$ where $\overline{\bm{z}}_{-1}$ is the vector of averages of all the moderator variables except $z_1$. As in Section \ref{sec:gb-sim}, we compared three methods: CMPBoost, CMPMOB  with split points estimated via exhaustive search, and CMPMOB  with split points estimated via 10\% change points. And similarly to the varying coefficients example in Section \ref{sec:gb-sim}, the results here illustrate that CMPBoost 
is able to reconstruct the underlying smooth function whereas CMPMOB provides only a piece-wise constant approximation. 

\begin{figure}
\centering
  \includegraphics[width=0.75\textwidth]{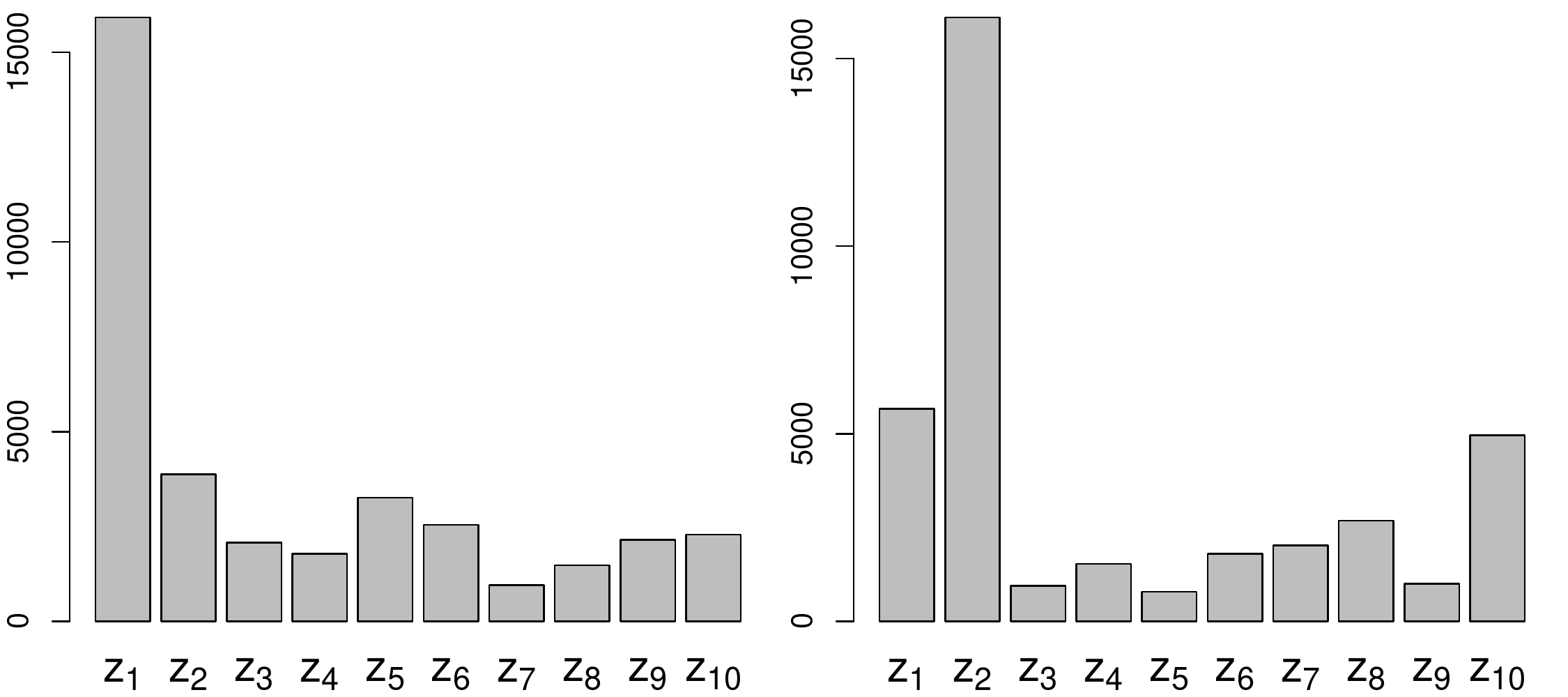}
\caption{Variable importance plots for the 10 moderator variables in CMPBoost using two different simulation datasets.}
\label{fig:boost-vi}
\end{figure}

Based on the results obtained from the CMPBoost model,  we  also constructed the
variable importance plot for the moderator variables 
(Figure \ref{fig:boost-vi}). As expected,   both $z_1$ and $z_2$
are the only significant moderator variables. For the majority of the datasets (out of 20 simulations), the variable importance measure is higher for  moderator variable $z_2$ although both $z_1$ and $z_2$ are used in the simulation. For a few simulations, the variable importance measure is higher for 
$z_1$. 
This might be a result of sampling variation.  

On the whole, these results reiterate the fact that
CMPBoost 
is a more flexible and robust approach than CMPMOB. 



%
\newpage
\section{Appendix A4: Bike Sharing Analysis - Comparison with Poisson and CMP GLM}
\label{apx: bik}

\begin{table}[!htbp] \centering 
\begin{tabular}{@{\extracolsep{5pt}}lccccccc} 
\\[-1.8ex]\hline 
\hline \\[-1.8ex] 
Statistic & \multicolumn{1}{c}{N} & \multicolumn{1}{c}{Mean} & \multicolumn{1}{c}{St. Dev.} & \multicolumn{1}{c}{Min} & \multicolumn{1}{c}{Pctl(25)} & \multicolumn{1}{c}{Pctl(75)} & \multicolumn{1}{c}{Max} \\ 
\hline \\[-1.8ex] 
holiday & 741 & 0.063 & 0.244 & 0 & 0 & 0 & 1 \\ 
weekday & 741 & 2.811 & 2.011 & 0 & 1 & 5 & 6 \\ 
workingday & 741 & 0.645 & 0.479 & 0 & 0 & 1 & 1 \\ 
temp & 741 & 0.275 & 0.103 & 0.020 & 0.200 & 0.340 & 0.580 \\ 
atemp & 741 & 0.275 & 0.101 & 0.015 & 0.212 & 0.333 & 0.546 \\ 
hum & 741 & 0.587 & 0.203 & 0.210 & 0.420 & 0.760 & 1.000 \\ 
windspeed & 741 & 0.217 & 0.130 & 0 & 0.1 & 0.3 & 1 \\ 
casual & 741 & 12.104 & 19.478 & 0 & 1 & 14 & 156 \\ 
registered & 741 & 118.455 & 109.922 & 1 & 29 & 168 & 518 \\ 
cnt & 741 & 130.559 & 119.797 & 1 & 32 & 191 & 559 \\ 
\hline \\[-1.8ex] 
\end{tabular}
\caption{Descriptive statistics for the bikesharing dataset. }
\label{tab:bike-descr}
\end{table} 

\begin{figure}[!htbp]
 \centering
  \includegraphics[height=0.5\textheight]{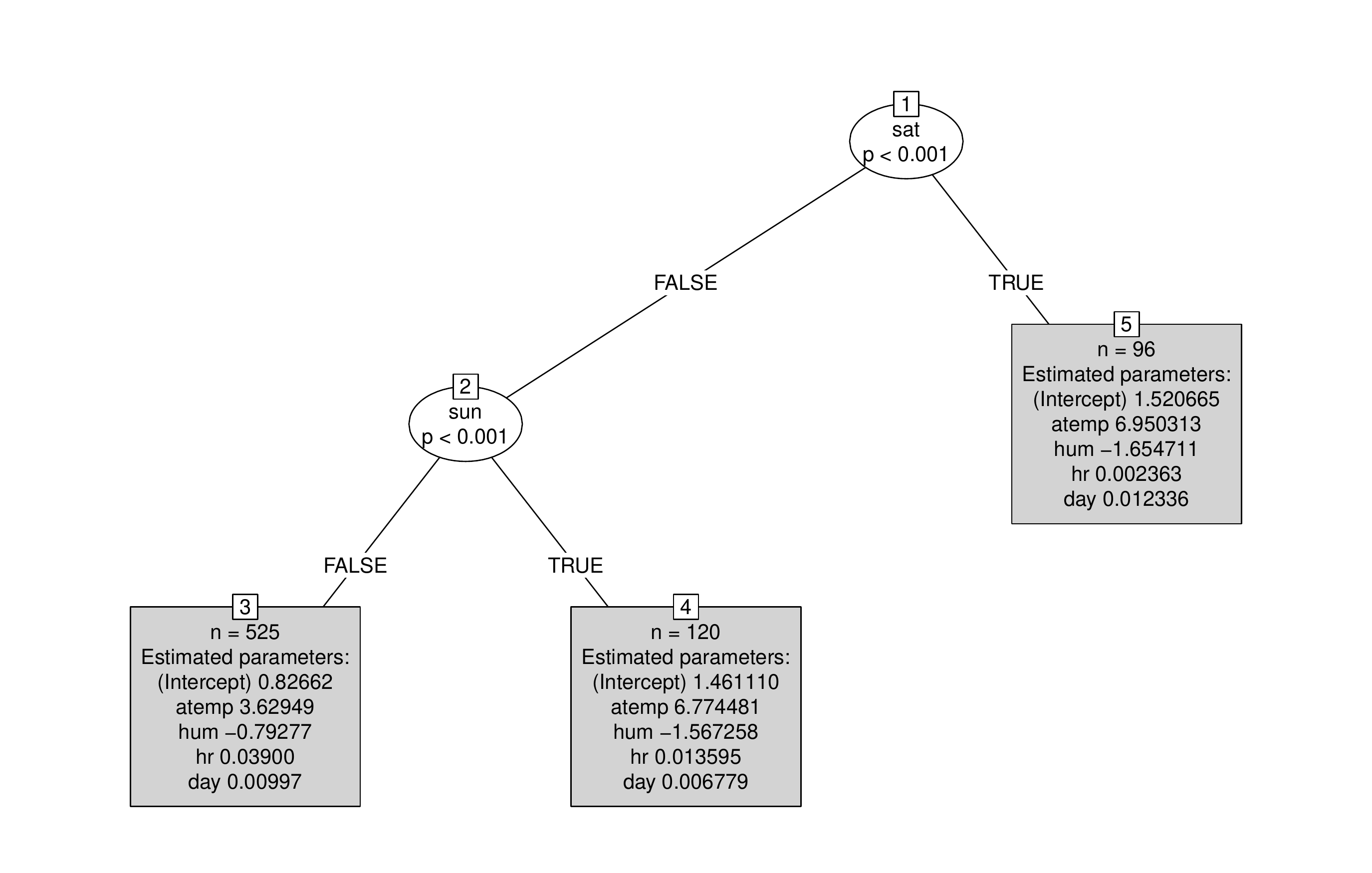}
  \caption{The Poisson MOB Tree for model
    (\ref{eqn:tvc}). The local likelihood value is $-4992.88$.}
  \label{fig:pos-mob}
\end{figure}

This appendix contains results that supplement Section \ref{sec:mob-bikeshare}. Table \ref{tab:glm-results} contains the results from both Poisson and CMP generalized linear models. Not surprisingly, the CMP GLM performed better in terms of overall fit. In addition, Poisson regression, which is not flexible enough to model the data over dispersion,   identifies \emph{hum} and \emph{windspeed} as significant  while the CMP GLM does not. Since the MOB framework uses coefficient constancy tests, it is possible that the limitations of Poisson regression are carried over to the Poisson MOB tree. For this reason, the Poisson MOB tree in Figure \ref{fig:pos-mob} looks completely different from the CMPMOB tree in Figure \ref{fig:cmp-mob}.

\begin{figure}[!htbp]
	\centering
	\includegraphics[width=\textwidth]{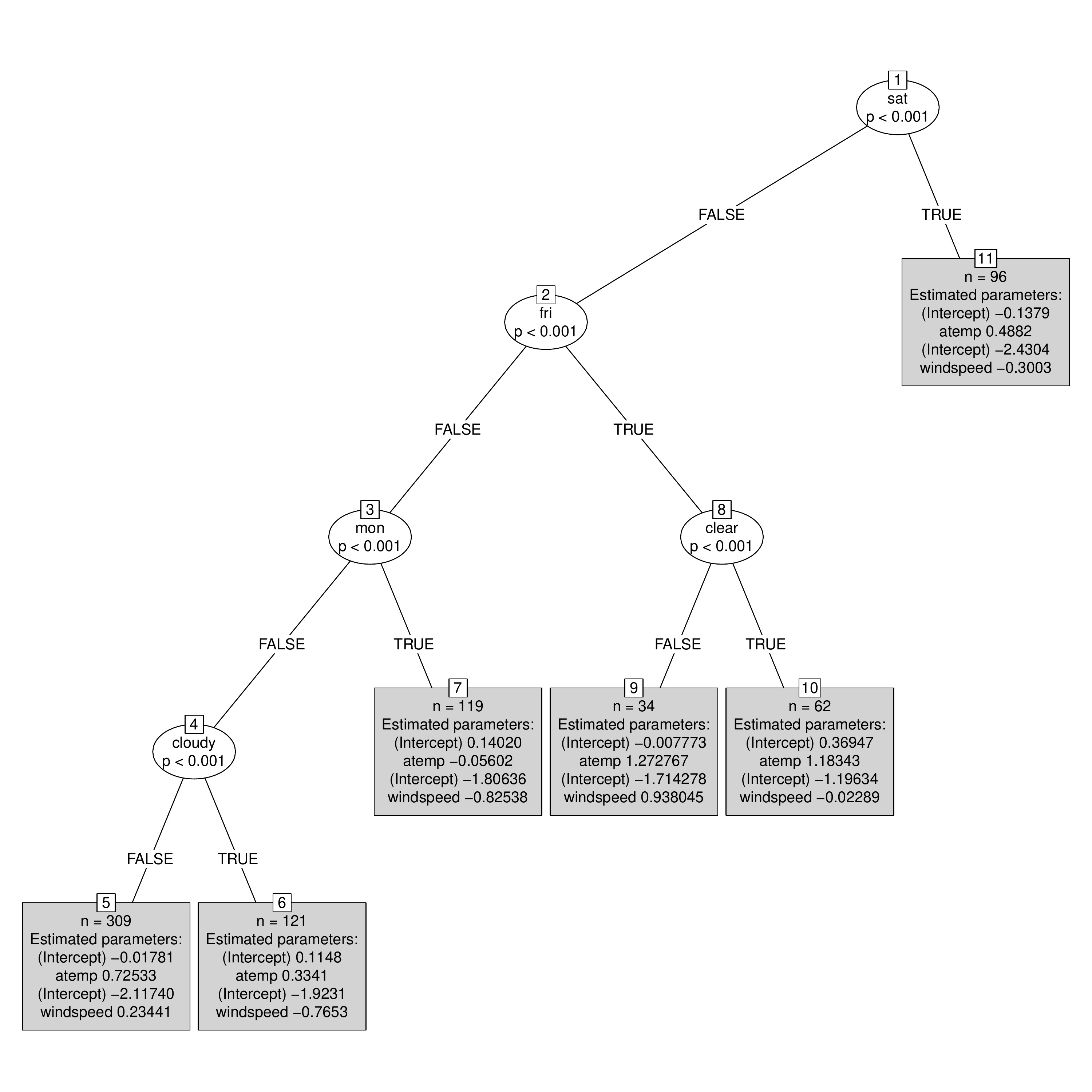}
	\caption{The tree for the  CMPMOB semi-varying coefficient model in
    (\ref{eqn:tvc-gen}). The likelihood for the tree model ($-1964.50$) is better than the model without any varying coefficient ($-2046.88$).}
	\label{fig:cmp-mob-gen}
\end{figure}

\begin{table}[!htbp] \centering 
\small
  \caption{Estimated regression coefficients and standard errors for
  the Poisson and CMP generalized linear models.} 
  \label{tab:glm-results} 
\begin{tabular}{lcc} 
\\[-1.8ex]\hline 
\hline \\[-1.8ex] 
  & \textit{Poisson GLM} & \textit{CMP GLM}\\ 
\hline \\[-1.8ex] 
 day & 0.007$^{***}$ (0.001) & 0.001$^{***}$ (0.001)\\ 
 hr & 0.026$^{***}$ (0.002) & 0.003$^{***}$ (0.001)\\ 
 holiday: holiday & 0.629$^{***}$ (0.075) & 0.113$^{**}$ (0.034) \\ 
 weekday: mon & $-$0.998$^{***}$ (0.058) & $-$0.138$^{***}$(0.028)\\ 
 weekday: tue & $-$1.100$^{***}$ (0.040) &$-$0.126$^{***}$ (0.015)\\ 
 weekday: wed & $-$0.641$^{***}$ (0.043) & $-$0.070$^{***}$ (0.017) \\ 
 weekday: thu & $-$0.919$^{***}$ (0.044) & $-$0.103$^{***}$ (0.018) \\ 
 weekday: fri & $-$1.021$^{***}$ (0.041) & $-$0.110$^{***}$ (0.016)\\ 
 weekday: sat & 0.021 (0.031) & $-$0.019$^{\cdot}$ (0.010)\\ 
 weathersit: cloudy & $-$0.165$^{***}$ (0.030) & $-$0.028$^{*}$ (0.011) \\ 
 weathersit: light rain/fog & $-$0.844$^{***}$ (0.070) & $-$0.145$^{***}$(0.037) \\ 
  weathersit: heavy rain/fog & $-$0.353  (0.380) & $-$0.101 (0.228) \\
  atemp & 5.709$^{***}$ (0.114) & 0.617$^{***}$ (0.047)\\ 
  hum & $-$0.595$^{***}$ (0.074) & $-$0.045 (0.029) \\ 
  windspeed & 0.326$^{***}$  (0.096) & 0.010 (0.036) \\ 
  Constant & 1.155$^{***}$  (0.069)  & $-$0.108$^{***}$ (0.028)\\
  \hline \\
  $\widehat{\gamma}_0$ & & $-$2.99$^{***}$ (0.022) \\
  \hline \\[-1.8ex] 
 Observations & 741 & 741\\ 
 Log Likelihood & $-$4951.48 & $-$2417.20  \\ 
 Akaike Inf. Crit. & 9934.96 & 4869.40  \\ 
 \hline 
\hline \\[-1.8ex] 
 & \multicolumn{1}{r}{$^{\cdot}$p$<$0.1; $^{*}$p$<$0.05; $^{**}$p$<$0.01}; $^{***}$p$<$0.001 \\ 
\end{tabular} 
\end{table} 
\begin{figure}[!htbp]
	\centering
	\includegraphics[scale=0.5]{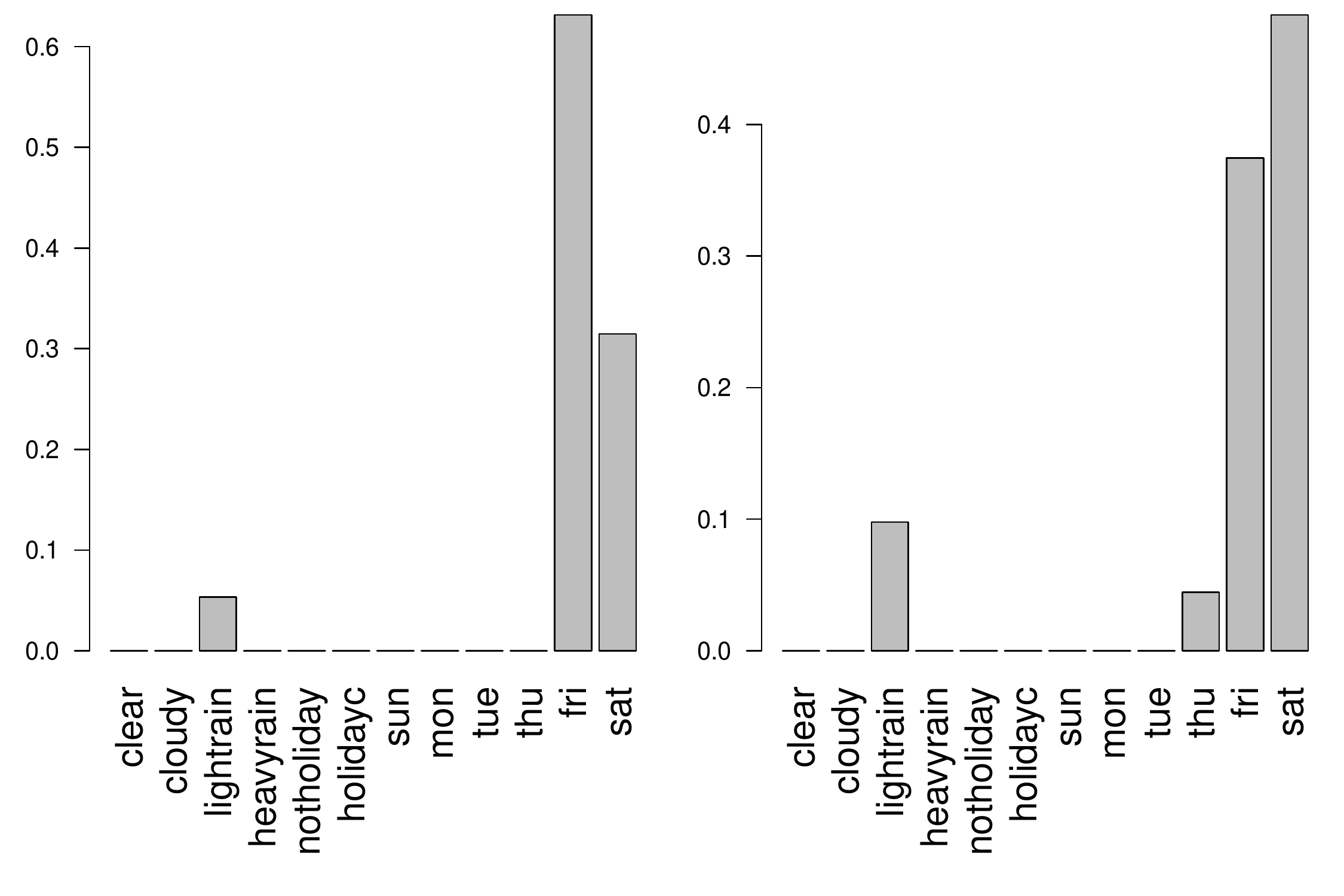}
    \caption{Variable importance plots for the CMPBoost model
      in (\ref{eqn:tvc-gen}). }
    \label{fig:boost-vi-gen}
\end{figure}
For the more flexible model in (\ref{eqn:tvc-gen}), the CMPMOB tree is given in Figure \ref{fig:cmp-mob-gen}. Similar to the CMPMOB tree in Figure \ref{fig:cmp-mob}, it identified weather and weekday as the significant factors affecting  demand after controlling for the other factors such as \emph{atemp} and \emph{hum}.

For the CMPBoost semi-varying coefficient model, the variable importance results for both the parameters are given in Figure \ref{fig:boost-vi-gen}. As mentioned in Section \ref{sec:mob-bikeshare}, while the $\ln \lambda$ model identified \emph{fri} and \emph{sat} as the top 2 variables, the $\ln \nu$ model identified \emph{sat} and \emph{fri} as the top 2 variables. These results illustrate the flexibility of the CMPBoost semi-varying coefficient model.

%

%

\end{document}